\documentclass[12pt]{article}

\usepackage{amssymb}
\usepackage{amsmath}
\usepackage{latexsym}

\catcode`@=11
\renewcommand{\section}{\@startsection{section}{1}{0pt}{\medskipamount}
{\medskipamount}{\large\bf}}
\numberwithin{equation}{section}
\catcode`@=12
\def\beq{\begin{eqnarray}}    %%%  begequation/eqnarray
\def\eeq{\end{eqnarray}}      %%%  endequation/eqnarray

\def\ln{\,\mbox{ln}\,}                  %%% log
\def\sTr{\,\mbox{sTr}\,}                %%% superTrace
\def\sDet{\,\mbox{sDet}\,}              %%% superdeterminant
\def\pa{\partial}                       %%% partial
\def\={\ =\ }

\begin{document}

\begin{center}

{\Large\bf Finite anticanonical  transformations in
field-antifield formalism}

\vspace{18mm}

{\Large Igor A. Batalin$^{(a, b)}\footnote{E-mail:
batalin@lpi.ru}$\;, Peter M. Lavrov$^{(b, c)}\footnote{E-mail:
lavrov@tspu.edu.ru}$\;, Igor V. Tyutin$^{(a, b)}\footnote{E-mail:
tyutin@lpi.ru}$ }

\vspace{8mm}

\noindent ${{}^{(a)}}$
{\em P.N. Lebedev Physical Institute,\\
Leninsky Prospect \ 53, 119 991 Moscow, Russia}

\noindent  ${{}^{(b)}}
${\em
Tomsk State Pedagogical University,\\
Kievskaya St.\ 60, 634061 Tomsk, Russia}

\noindent  ${{}^{(c)}}
${\em
National Research Tomsk State  University,\\
Lenin Av.\ 36, 634050 Tomsk, Russia}

\vspace{20mm}

\begin{abstract}
\noindent We study the role of arbitrary (finite) anticanonical
transformations in the field-antifield formalism and the
gauge-fixing procedure based on the use of these transformations.
 The properties of the generating functionals of the Green functions
subjected to finite anticanonical transformations are considered.
\end{abstract}

\end{center}

\vfill



\newpage

\section{Introduction}

The field-antifield formalism \cite{BV,BV1}, summarizing numerous
attempts to find correct quantization rules for various types of
gauge models \cite{dWvH,FT,Nie1,Kal,Nie2}, is a powerful covariant
quantization method which can be applied to arbitrary gauge
invariant systems. This method is based on the fundamental principle
of BRST invariance \cite{brs1,t} and has a rich new geometry
\cite{Sch}. One of the most important objects of the field-antifield
formalism is an odd symplectic structure called antibracket and
known to mathematicians  as the Buttin bracket \cite{Butt}. In terms
of the antibracket the master equation and the Ward identity for
generating functional of the vertex functions (effective action) are
formulated. It is an important property that the antibracket is
preserved under the anticanonical transformations which are dual to
canonical transformations for a Poisson bracket. An important role
and rich geometric possibilities of general anticanonical
transformations in the field-antifield formalism have been realized
in the procedure of gauge fixing \cite{VLT} (see, also \cite{BBD}).
The original procedure of gauge fixing \cite{BV,BV1} corresponds in
fact to a special type of anticanonical transformation in an action
being a proper solution to the quantum master equation. That type of
transformations is capable to yield admissible gauge-fixing
conditions in the form of equations of arbitrary Lagrangian surfaces
(constraints in the antibracket involution) in the field-antifield
phase space. Thereby, the necessary class of admissible gauges was
involved actually. The latter made it possible to describe in
\cite{VLT} the structure and renormalization of general gauge
theories in terms of anticanonical transformations. As the authors
\cite{VLT} assumed the use of regularizations   in which $\delta(0)
= 0$ in local field theories, they based themselves on the use of
general anticanonical transformations in an action being a proper
solution to the classical master equation. In turn, the gauge
dependence and the structure of renormalization of the effective
action have been analyzed by using  infinitesimal anticanonical
transformations only.

In the present article, we extend the use of anticanonical
transformations in the field-antifield formalism from the
infinitesimal level to the finite one, and explore a gauge fixing
procedure for general  gauge theories, based on arbitrary
anticanonical transformations in an action  being a proper solution
to the quantum master equation with fixed boundary condition. Now it
is worthy to notice the difference between the properties of the
classical and quantum master equations under anticanonical
transformations. The classical master equation is covariant under
anticanonical transformations, as its left-hand side is the
antibracket of the action with itself. In contrast to that, the form
of the quantum master equation is not maintained under anticanonical
transformations. One should accompany the anticanonical
transformation by multiplying the exponential of $i/\hbar$ times the
transformed action with the square root of the superjacobian of that
anticanonical transformation. We will call such an operation  an
{\it anticanonical master transformation} and the corresponding
action a {\it master transformed action}. Thus, one can say that
the form of the quantum master equation is maintained under the
anticanonical master transformation.

We consider in all details the relationship between
the two descriptions (in terms of the generating functions and the generators)
for arbitrary finite anticanonical transformations.

Finally, let us notice the study \cite{BLT-BV}, among the other
recent developments, where a procedure was found to connect
generating functionals of the Green functions for a gauge system
formulated  in any two admissible gauges with the help of finite
field-dependent BRST transformations.
\\

\section{Field-Antifield Formalism}

\noindent

The starting point of the field-antifield formalism \cite{BV} is a
theory of fields $\{{\cal A}\}$ for which the initial classical
action $S_0({\cal A})$ is assumed to be invariant under the gauge
transformations $\delta {\cal A}=R({\cal A}) \xi$. Here $\xi$ are
arbitrary functions of space-time coordinates , and $\{R({\cal
A})\}$ are generators of gauge transformations. The set of
generators is complete but, in general, maybe reducible and forms an
open gauge algebra so that one works with general gauge theories.
Here we do not discuss these points, referring to original papers
\cite{BV,BV1}. The structure of the gauge algebra determines
necessary content of total configuration space of fields
$\{\varphi^i\;(\varepsilon(\varphi^i)=\varepsilon_i)\}$ involving
fields
 $\{{\cal A}\}$ of initial
classical system, ghost and antighost fields, auxiliary fields and ,
in case of reducible generators, pyramids of extra ghost and
antighost fields as well as pyramids of extra auxiliary fields.
To each field $\varphi^i$  one introduces an antifield
$\varphi^*_i$, whose statistics is opposite to that of the
corresponding fields $\varphi^i$,  $\varepsilon (\varphi^*_i) =
\varepsilon_i + 1 $. On the space of the fields $\varphi^i$ and
antifields $\varphi^*_i$ one defines an odd symplectic structure
$(\;,\;)$ called the antibracket
\begin{eqnarray}
\label{defAB} (F, G)\equiv
F\left(\overleftarrow{\pa}_{\varphi^i}\overrightarrow{\pa}_{\varphi^*_i}-
\overleftarrow{\pa}_{\varphi^*_i}\overrightarrow{\pa}_{\varphi^i}\right)G
\end{eqnarray}
and the nilpotent fermionic operator $\Delta$,
\begin{eqnarray}
\label{DeltaBV} \Delta=(-1)^{\varepsilon_i} \partial_{\varphi^{i}}\partial
_{\varphi_{i}^{\ast}},\quad\Delta^{2}=0,
\quad \varepsilon (\Delta)=1.
\end{eqnarray}
Here the notation
\beq
\pa_{\varphi^i}=\frac{\pa }{\pa\varphi^i},\quad
\pa_{\varphi^*_i}=\frac{\pa}{\pa\varphi^*_i}
\eeq
is introduced. In terms of the antibracket and
$\Delta$-operator the quantum master equation is formulated
as
\begin{eqnarray}
\label{MastEBV} \frac {1}{2}
({\cal S},{\cal S})=i\hbar{\Delta}{\cal S}\qquad\Leftrightarrow\qquad
\Delta\;\exp\Big\{\frac{i}{\hbar}{\cal S}\Big\}=0
\end{eqnarray}
for a bosonic functional
${\cal S}={\cal S}(\varphi,\varphi^*)$ satisfying the
boundary condition
\begin{eqnarray}
\label{BoundCon} {\cal S}|_{\varphi^* = \hbar = 0}= S_0({\cal A})
\end{eqnarray}
and being the basic object of the field-antifield quantization
scheme \cite{BV,BV1}. Among the properties of the antibracket and
$\Delta$-operator we mention the Leibniz rule,
\beq
(F, GH)=(F,G)H+(F, H)G(-1)^{\varepsilon(G)\varepsilon(H)},
\eeq
the Jacobi identity,
\beq
\label{JI}
((F, G),
H)(-1)^{(\varepsilon(F)+1)(\varepsilon(H)+1)} +{\sf cycle} (F, G,
H)\equiv 0,
\eeq
and the $\Delta$-operator being a derivative to the
antibracket,
\beq
\label{DAnt}
\Delta (F,G)=(\Delta F,G)-(F,\Delta
G)(-1)^{\varepsilon(F)}.
\eeq
There exists a generating functional
$Y=Y(\varphi,\Phi^*),\;\varepsilon(Y)=1$ of the anticanonical
transformation, \beq \label{antican}
\Phi^i=\pa_{\Phi^*_i}Y(\varphi,\Phi^*),\quad \varphi^*_i=
Y(\varphi,\Phi^*)\overleftarrow{\pa}_{\varphi^i}. \eeq The
invariance property of the odd symplectic structure (\ref{defAB}) on
the phase space of $(\varphi,\varphi^*)$ is dual to the invariance
property of an even symplectic  structure (a Poisson bracket) under
a canonical transformation of canonical variables $(p,q)$ (for
further discussions of relations between Poisson bracket and
antibracket, see \cite{BH,BM}).

The generating functional of the Green functions $Z(J)$ is defined in
terms of the functional integral as \cite{BV,BV1}
\begin{eqnarray}
\label{ZBV} {\cal Z}(J)=\int {\cal D}\varphi\;
\exp\left\{\frac{i}{\hbar}\left[S_{e}(\varphi)+ J_i\varphi^i\right]\right\}
=\exp\left\{\frac{i}{\hbar}W(J)\right\},
\end{eqnarray}
where
\beq
S_{e}(\varphi) = {\cal S}\big(\varphi,\varphi^*=\pa_{\varphi}\psi(\varphi)\big),
\eeq
 $\psi(\varphi)$ is a fermionic gauge
 functional,  $J_i$
$(\varepsilon(J_i) = \varepsilon_i)$ are  usual external sources
to the fields $\varphi^i$ and $W(J)$ is the generating functional
of the connected Green functions.

 To discuss the quantum properties of general
gauge theories, it is useful to consider, instead of the generating functional
(\ref{ZBV}), the extended generating functionals
${\cal Z}(J,\varphi^*)$ and $W(J,\varphi^*)$
defined by the relations
\beq \label{ZBVEx}
{\cal Z}(J,\varphi^*)=\int {\cal D}\varphi\;
\exp\Big\{\frac{i}{\hbar}\big[S(\varphi,\varphi^*)+J_i\varphi^i\big]\Big\}=
\exp\left\{\frac{i}{\hbar}W(J,\varphi^*)\right\}
\eeq
where
\beq
\label{AcBV}
S(\varphi,\varphi^*)={\cal S}\big(\varphi,\varphi^*+
\pa_{\varphi}\psi(\varphi)\big).
\eeq
Obviously, we have
\beq {\cal Z}(J)={\cal Z}(J,\varphi^*)\mid_{\varphi^*=0},
\quad W(J)=W(J,\varphi^*)\mid_{\varphi^*=0}.
\eeq
The action $S=S(\varphi,\varphi^*)$
(\ref{AcBV}) satisfies the quantum master equation
\beq
\label{MEext} \frac{1}{2}(S,S)=i\hbar \Delta
S\qquad\Leftrightarrow\qquad \Delta
\exp\Big\{\frac{i}{\hbar}S\Big\}=0.
\eeq
It follows from (\ref{MEext}) that the Ward identities hold  for the extended
generating functionals  ${\cal Z}(J,\varphi^*)$ and $W(J,\varphi^*)$
\beq \label{WIZ}
J_i\pa_{\varphi^*_i}{\cal Z}(J,\varphi^*)=0\;,\quad
J_i\pa_{\varphi^*_i}W(J,\varphi^*)=0.
\eeq
Indeed, we have
\beq
\nonumber
&&0=\int d\varphi \exp\left\{\frac{i}{\hbar}J\varphi\right\}
\left(\Delta \exp\left\{\frac{i}
{\hbar}S\right\}\right)=\int d\varphi(-1)^{\varepsilon_i}
\partial_{\varphi^{i}}\left[
\exp\left\{\frac{i}{\hbar}J\varphi\right\}\partial_{\varphi_{i}^{\ast}}
\exp\left\{\frac{i}{\hbar}
S\right\}\right]  -\\
\nonumber
&&-\frac{i}{\hbar}J_{i}\partial_{\varphi_{i}^{\ast}}\int d\varphi
\exp\left\{\frac{i}{\hbar}(S+J\varphi)\right\}=
-\frac{i}{\hbar}J_{i}\partial_{\varphi
_{i}^{\ast}}{\cal Z}(J,\varphi^*)=\\
\label{WIW}
&&=-\frac{i}{\hbar}J_{i}\partial_{\varphi_{i}^{\ast}}
\exp\left\{\frac{i}{\hbar}W(\varphi^{\ast},J)\right\}
\Longrightarrow
J_{i}\partial_{\varphi_{i}^{\ast}}W(\varphi^{\ast},J)=0.
\eeq

 The generating functional of vertex function
(effective action) is defined via the Legendre transformation
\beq
\Gamma(\varphi,\varphi^*)=W(J,\varphi^*)-J\varphi,\;
\varphi^i=\pa_{J_i} W(J,\varphi^*),\;
\partial_{\varphi_{i}^{\ast}}W(J,\varphi^{\ast})=
\partial_{\varphi_{i}^{\ast}}\Gamma(\varphi,\varphi
^{\ast}),\; \pa_{J_i}=\frac{\pa}{\pa J_i}
\eeq
with the properties
\beq
\label{prG}
J_{i}=-\Gamma(\varphi,\varphi^{\ast})\overleftarrow{\partial}_{\varphi^{i}
}=-(-1)^{\varepsilon_i}\Gamma_{i},\;\Gamma_{i}=\Gamma_{i}(\varphi,\varphi^{\ast}
)=\partial_{\varphi^{i}}\Gamma(\varphi,\varphi^{\ast})\:.
\eeq
It follows from (\ref{WIW}) and (\ref{prG}) that the Ward
identity for the effective action holds,
\beq
\label{WIG}
\Gamma\overleftarrow{\partial}_{\varphi^{i}}\partial_{\varphi_{i}^{\ast}%
}\Gamma=0\;\Longrightarrow\frac{1}{2}(\Gamma,\Gamma)=0,
\eeq
which has the form of classical master equation in the field-antifield formalism.

As it was pointed out for the first time in \cite{VLT}, the gauge
fixing procedure in the field-antifield formalism
(\ref{AcBV}) can be described in terms of a special type
of anticanonical transformation (\ref{antican}).
Indeed, let us consider anticanonical transformations of the variables
($\varphi,\varphi^*$) with specific generating function
\beq
\label{santican}
Y=Y(\varphi,\Phi^*)=\Phi^*_i\varphi^i-\psi(\varphi).
\eeq
We have
\beq
\Phi^i=\varphi^i,\quad \varphi^*_i=\Phi^*_i-\pa_{\varphi^i}\psi(\varphi),
\eeq
so that the transformed action ${\tilde S}={\tilde S}(\varphi,\varphi^*)$
\beq
{\tilde S}(\varphi,\varphi^*)=S(\Phi,\Phi^*)=S(\varphi,\varphi^*+
\pa_{\varphi}\psi(\varphi))
\eeq
coincides with (\ref{AcBV}). In particular, this fact made it possible
to study effectively the
gauge dependence and structure of renormalization of general
 gauge theories \cite{VLT}.
In what follows we explore a gauge fixing procedure in the field-antifield
formalism as an
anticanonical transformation of general type with the only
requirement for anticanonically
generalized action:
the supermatrix of the second field derivatives of this action must be non-degenerate.
An essential difference in this point with the approach used in \cite{VLT} is
that we work with a general setting for an action (\ref{AcBV}) which
satisfies the quantum master equation (not the classical master
equation as in \cite{VLT}).
\\

\section{Infinitesimal anticanonical transformations}

As the first step in our study of anticanonical transformations in
the field-antifield formalism, we consider the properties of the main
objects subjected to infinitesimal anticanonical transformations. In
the latter case, the generating functional $Y$ reads
\beq
\label{infantican}
Y=Y(\varphi,\Phi^*)=\Phi^*_i\varphi^i+X(\varphi,\Phi^*), \quad
\varepsilon(X)=1.
\eeq
The functional $X$ is considered as the
infinitesimal one. Then the anticanonical transformations of the variables,
\beq \Phi^i=\varphi^i+\pa_{\Phi^*_i}X(\varphi,\Phi^*),\quad
\varphi^*_i=\Phi^*_i+\pa_{\varphi^i}X(\varphi,\Phi^*),
\eeq
can be written down to the first order in $X$ as
\beq
\Phi^i=\varphi^i+\pa_{\varphi^*_i}X(\varphi,\varphi^*)+O(X^2),\quad
\Phi^*_i=\varphi^*_i-\pa_{\varphi^i}X(\varphi,\varphi^*)+O(X^2)
\eeq
or, in terms of the antibracket (\ref{defAB}),
\beq
\Phi^i=\varphi^i+(\varphi^i,X)+O(X^2),\quad
\Phi^*_i=\varphi^*_i+(\varphi^*_i,X)+O(X^2). \eeq The
anticanonically transformed action ${\tilde S}$,
\beq
\label{trnS}
{\tilde S}={\tilde
S}(\varphi,\varphi^*)=S(\Phi,\Phi^*)=S+(S,X)+O(X^2)
\eeq
does not
satisfy the quantum master equation to the first approximation in $X$,
\beq
\frac{1}{2}({\tilde S},{\tilde S})-i\hbar \Delta {\tilde
S}=i\hbar(S,\Delta X)+O(X^2)\neq 0.
\eeq
Consider now the
superdeterminant of the anticanonical transformation
\beq
\label{Jacob} {\cal J}(\varphi,\varphi^*)={\cal J}(Z)=\sDet
\left[{\bar Z}^A(Z) \overleftarrow{\pa}_B\right],
\eeq
where
\beq
{\bar Z}^A=(\Phi^i,\Phi^*_i),\quad Z^A=(\varphi^i,\varphi^*_i),\quad
\pa_A=\frac{\pa}{\pa Z^A}.
\eeq
To the first-order approximation in
$X$, the ${\cal J}$ reads
\beq
{\cal J}=\exp\{2\Delta X\}+O(X^2)=
\exp\left\{\frac{i}{\hbar}\big(-2i\hbar\Delta X\big)\right\}+O(X^2).
\eeq
In contrast to the notation used in  \cite{LT,BV2}, now we refer to
$S'=S'(\varphi,\varphi^*)$ constructed from $S=S(\varphi,\varphi^*)$
via the {\it anticanonical master transformation},
\beq
\label{E3.10}
S'&=&S'(\varphi,\varphi^*)=
S(\Phi(\varphi,\varphi^*),\Phi^*(\varphi,\varphi^*))-
i\hbar\frac{1}{2}\ln {\cal J}(\varphi,\varphi^*), \label{trAc}
\eeq
as the {\it master-transformed action}.

Note that, by itself, the anticanonical master transformation can be defined
without reference  on
solutions of the quantum master equation. Namely, let us define a transformation
of the form\footnote{In the present article, we only consider the case
in which the generator $F$ is a function; the case of
operator-valued $F$ it was studied in \cite{BB1}.}
\beq
\exp\left\{\frac{i}{\hbar}G'\right\}=
\exp\{-[F,\Delta]\}\exp\left\{\frac{i}{\hbar}G\right\},
\eeq
where $G, F$ ($\varepsilon(G)=0,\;\varepsilon(F)=1$)
are  arbitrary functions of $\varphi,\varphi^*$, and $[\;,\;]$
stands for the supercommutator.
Then we can prove
(see Appendices C and D) the relation
\beq
\label{E3.12}
G'=\exp\{{\rm ad}(F)\}G+i\hbar f({\rm ad}(F))\Delta F,
\; f({\rm ad}(F))\Delta F=-\frac{1}{2}\ln {\cal J},\; {\rm ad}(F)(...)=(F, (...)),
\eeq
which repeats the relation (\ref{E3.10}). In (\ref{E3.12}) the notation
$f(x)=(\exp x -1)x^{-1}$ is used.

The action $S'$ (\ref{trAc})
to the first order in $X$
\beq
S'=S+(S,X)-i\hbar\Delta X+O(X^2)
\eeq
does satisfy the quantum master equation
\beq \label{QMEStr}
\frac{1}{2}(S',S')-i\hbar\Delta S'=O(X^2).
\eeq
Note that, due to the
results of \cite{LT,BV2}, the action (\ref{trAc}) by itself
satisfies the quantum master equation in the case of arbitrary
anticanonical transformation, as well (see, also \cite{BLT4,BBD}).

Let us consider the generating functionals constructed
with the help of master-transformed action $S^{\prime}$ to the first order
in $X$. We have
\beq
\nonumber &&{\cal Z}'
={\cal Z}'(J,\varphi^{\ast})=\int d\varphi
\exp\left\{\frac{i}{\hbar}(S'+J\varphi)\right\}=
\exp\left\{\frac{i}{\hbar}W^{\prime}(J,\varphi^{\ast})\right\}=\\
&&=\exp\left\{\frac{i}{\hbar}W(J,\varphi^{\ast})
\right\}\left(1+\frac{i}{\hbar}\delta
W(J,\varphi^{\ast})\right)\!,\\
&&\Gamma'(\varphi,\varphi^{\ast})  =W'(J,\varphi^{\ast})
-J\varphi=\Gamma(\varphi,\varphi^{\ast})+\delta\Gamma(\varphi
,\varphi^{\ast}),\\
\nonumber
&&\delta\Gamma(\varphi,\varphi^{\ast})=
\delta W(J(\varphi,\varphi^{\ast}),\varphi^*).
\eeq
Therefore
\beq
\nonumber
 &&{\cal Z}'-{\cal Z}=\delta {\cal Z}=
 \frac{i}{\hbar}\exp\left\{\frac{i}{\hbar}W(J,\varphi^{\ast})\right\}
\delta
W(J,\varphi^{\ast})=\\
\nonumber
&&=\frac{i}{\hbar}\exp\left\{\frac{i}{\hbar}W(J,\varphi^{\ast})\right\}
\delta\Gamma(\varphi,\varphi^{\ast})=\nonumber\\
\nonumber
&&=\frac{i}{\hbar}\int d\varphi\left[  (S,X)-i\hbar\Delta X\right]
\exp\left\{\frac{i}{\hbar}(S+J\varphi)\right\}=\\
\nonumber
&&=\int d\varphi \exp\left\{\frac{i}{\hbar}J\varphi
\right\}\Delta\left(  X\exp\left\{\frac{i}{\hbar}S\right\}\right)  =\\
\nonumber
&& =-\frac{i}{\hbar}J_{i}\partial_{\varphi_{i}^{\ast}}\left[  \tilde
{X}(J,\varphi^{\ast})\exp\left\{\frac{i}{\hbar}W(J,\varphi^{\ast})\right\}\right]=\\
\label{delZ}
&&=\exp\left\{\frac{i}{\hbar}W(J,\varphi^{\ast})\right\}\left[-\frac{i}{\hbar}J_{i}
\partial_{\varphi_{i}^{\ast}}\tilde{X}(J,\varphi^{\ast})\right],
\eeq
where
\beq
\tilde{X}(J,\varphi^{\ast})=
\exp\left\{-\frac{i}{\hbar}W(J,\varphi^{\ast})\right\}\int
d\varphi X\exp\left\{\frac{i}{\hbar}(S+J\varphi)\right\}.
\eeq
When deriving (\ref{delZ}), the Ward identity for $W(J,\varphi^*)$
(\ref{WIW}), the quantum master equation for $S(\varphi,\varphi^*)$
(\ref{MEext}) and the following identities:
\beq
\label{usID}
i\hbar\exp\left\{\frac{i}{\hbar}S\right\}\Delta X=
i\hbar\Delta\left(  X\exp\left\{\frac{i}{\hbar} S\right\}\right)
+(S,X)\exp\left\{\frac{i}{\hbar}S\right\}, \eeq \beq \nonumber
&&\exp\left\{\frac{i}{\hbar}J\varphi\right\}
\Delta\left(X\exp\left\{\frac{i}{\hbar}S\right\}\right)
=(-1)^{\varepsilon_i}\partial_{\varphi^{i}}
\left[\exp\left\{\frac{i}{\hbar}J\varphi\right\}
\partial_{\varphi_{i}^{\ast}}\left(  Xe^{\frac{i}{\hbar}S}\right)  \right]-\\
\label{usID1}
&&\qquad-\frac{i}{\hbar}J_{i}\partial_{\varphi_{i}^{\ast}}
\left(  X\exp\left\{\frac{i}{\hbar
}(S+J\varphi)\right\}\right)
\eeq
are used. Rewriting (\ref{delZ}) for a variation of the effective action
$\Gamma=\Gamma(\varphi,\varphi^*)$, we obtain
\beq
\nonumber
&&\delta\Gamma(\varphi,\varphi^{\ast})=-J_{i}\partial_{\varphi_{i}^{\ast}}
\tilde{X}(J,\varphi^{\ast})=(-1)^{\varepsilon_i}
\Gamma_{i}\partial_{\varphi_{i}^{\ast}}
\mathcal{X}(\varphi,\varphi^{\ast})-\\
\label{delGamma}
&&\qquad  -(-1)^{\varepsilon_i}\Gamma_{i}\left[\partial_{\varphi_{i}^{\ast}}J_{j}
(\varphi,\varphi^{\ast})\right]  \left.  \partial_{J_{j}}\tilde{X}
(J,\varphi^{\ast})\right|  _{J=J(\varphi,\varphi^{\ast})},
\eeq
where
\beq
\mathcal{X}(\varphi,\varphi^{\ast})=
\left.\tilde{X}(J,\varphi^{\ast})\right|_{J=J(\varphi,\varphi^{\ast})}.
\eeq
One can rewrite Eq. (\ref{delGamma}) in terms of
$\Gamma=\Gamma(\varphi,\varphi^*)$ as
\begin{equation}
\delta\Gamma(\varphi,\varphi^{\ast})=\Gamma\overleftarrow{\partial}
_{\varphi^{i}}\partial_{\varphi_{i}^{\ast}}\mathcal{X}-\Gamma\overleftarrow
{\partial}_{\varphi_{i}^{\ast}}\partial_{\varphi^{i}}\mathcal{X}
=(\Gamma,\mathcal{X})=-(\mathcal{X},\Gamma). \label{anticGamma}
\end{equation}
This result is proved in Appendix A.
Equation (\ref{anticGamma}) means that any infinitesimal anticanonical master
transformation of the action $S$ (\ref{trnS})
with a generating functional $X$ induces an infinitesimal anticanonical
transformation in
 the effective action $\Gamma$ (\ref{anticGamma}) with a generating functional
 $\mathcal{X}$, provided
the generating functional of the Green functions is constructed via
the master-transformed action. An important goal of our present study
is a generalization
of this fact (for the first time known among results of paper \cite{VLT})
to the case of arbitrary (finite) anticanonical transformation.
\\

\section{Finite anticanonical transformation}

Consider an arbitrary (finite) anticanonical transformation described
by a generating functional $Y=Y(\varphi,\Phi^*),\;\varepsilon(Y)=1$,
\footnote{Note that any anticanonical transformation can be described
in terms of a generating functional.}
\beq
\label{acY}
\varphi_{i}^{\ast}=Y(\varphi,\Phi^{\ast})\overleftarrow{\partial}_{\varphi^i}
,\quad\Phi^{A}=\pa_{\Phi^{\ast}_i}Y(\varphi,\Phi^{\ast}).
\eeq
Let $Y$ have the form
\beq
\label{Ya}
Y(\varphi,\Phi^*)=\Phi_{i}^{\ast}\varphi^{i}+a f(\varphi
,\Phi^{\ast}), \quad \varepsilon(f(\varphi,\Phi^{\ast}))=1,
\eeq
where $a$ is a parameter. Then solution of equations (\ref{acY})
 up to second order in $a$ can be written as
\beq
\label{Phia}
&&\Phi^{i}=\varphi^{i}+a f\overleftarrow{\partial}_{\varphi^{\ast}_i}-
a^{2}(-1)^{(\varepsilon_{i}+1)(\varepsilon_{j}+1)}f\overleftarrow
{\partial}_{\varphi^{\ast}_j}
\overleftarrow{\partial}_{\varphi^{\ast}_i}\overrightarrow
{\partial}_{\varphi^j}f+O(a^{3}),\\
\label{Phi*a} &&
\Phi_{i}^{\ast}=\varphi_{i}^{\ast}-a\partial_{\varphi^i}f+a^{2}
(-1)^{\varepsilon_{i}(\varepsilon_{j}+1)}
f\overleftarrow{\partial}_{\varphi^{\ast}_j}
\overleftarrow{\partial}_{\varphi^i}
\overrightarrow{\partial}_{\varphi^j}f+O(a^{3}),
\eeq
where $f\equiv f(\varphi,\varphi^{\ast})$. Let us denote
\beq
Z^{A}=\{\varphi^{i},\varphi_{i}^{\ast}\},\quad
{\bar Z}^{A}=\{\Phi^{i},\Phi_{i}^{\ast}\},\quad
\varepsilon({\bar Z}^A)=\varepsilon(Z^A)= \varepsilon_A,
\eeq
 and
 \beq
\label{Fsc2}
F=F(\varphi,\varphi^{\ast};a)=-f(\varphi,\varphi^{\ast})+\frac{a}
{2}f(\varphi,\varphi^{\ast})\overleftarrow{\partial}_{\varphi^{\ast}_j}
\overrightarrow{\partial}_{\varphi^j}f(\varphi,\varphi^{\ast}).
\eeq
Then we have
\beq
\label{ZFa}
{\bar Z}^{A}={\bar Z}^{A}(Z;a)=\exp\{a{\rm ad}(F)\}Z^{A}+O(a^{3}),
\eeq
where ${\rm ad}(F)$ means the left adjoint of the antibracket
\beq \label{Fhat}
{\rm ad}(F)(...)=(F(Z;a),(...)).
\eeq
We call $F$ in (\ref{Fsc2}) a
generator of the anticanonical transformation to the second order. It
should be noticed that
the generator of an anticanonical transformation does
not coincide with the generating functional of this transformation
already to the second order.
A natural question arises: Does a generator  exist for a given anticanonical
transformation, actually? To answer this question,
we begin with the claim
that an operator $\exp\{{\rm ad} (F)\}$ generates
anticanonical transformation. Indeed, let $Z^A$ be anticanonical
variables so that the antibracket (\ref{defAB}) can be presented in
the form
\beq
\label{ABz}
(H(Z),G(Z))=H(Z)\overleftarrow{\partial}_{A}E^{AB}\overrightarrow{\partial}
_{B}G(Z),\quad
(Z^{A},Z^{B})=E^{AB},\quad\partial_{A}=\frac{\partial}{\partial Z^{A}},
\eeq
where $E^{AB}$ is a constant supermatrix with the properties
\beq
E^{BA}=-(-1)^{(\varepsilon_A+1)(\varepsilon_B+1)}E^{AB},\quad
\varepsilon(E^{AB})=\varepsilon_A+\varepsilon_B+1.
\eeq
Then the transformation
\beq
\label{trzZ}
Z^{A}\rightarrow
{\bar Z}^{A}(Z)=\exp\{{\rm ad}F(Z)\}Z^{A}
\eeq
is anticanonical,
\beq
({\bar Z}^{A}(Z),{\bar Z}^{B}(Z))={\bar Z}^{A}(Z)\overleftarrow{\partial}_{C}
E^{CD}\overrightarrow
{\partial}_{D}{\bar Z}^{B}(Z)=E^{AB}.
\eeq
To prove this fact we introduce
an one-parameter family of transformations
\beq
\label{Zi}
{\bar Z}^A(Z,a)=\exp\{a{\rm ad}( F)\}Z^A, \quad {\bar Z}^A(Z,0)=Z^A,
\eeq
and quantities ${\bar Z}^{AB}(Z,a)$,
\beq
\label{Zij}
{\bar Z}^{AB}(Z,a)=({\bar Z}^A(Z,a),{\bar Z}^B(Z,a)),
\quad {\bar Z}^{AB}(Z,0)=E^{AB}.
\eeq
It follows from the definitions (\ref{Zi}) and (\ref{Zij}), that the
relations
\beq
&&\frac{d}{da}{\bar Z}^A(Z,a)=(F(Z),{\bar Z}^A(Z,a)),\\
\nonumber
&&\frac{d}{da}{\bar Z}^{AB}(Z,a)=((F(Z),{\bar Z}^A(Z,a)),{\bar Z}^B(Z,a))
+({\bar Z}^A(Z,a),(F(Z),{\bar Z}^B(Z,a))=\\
\label{eqZij}
&&=(F(Z),({\bar Z}^A(Z,a),{\bar Z}^B(Z,a))=
(F(Z),{\bar Z}^{AB}(Z,a))={\rm ad}( F(Z)){\bar Z}^{AB}(Z,a),
\eeq
hold,
where the Jacobi identity (\ref{JI}) for antibrackets is used.
A solution to Eq. (\ref{eqZij}) has the form
\beq
{\bar Z}^{AB}(Z,a)=\exp\{a{\rm ad}( F(Z))\}{\bar Z}^{AB}(Z,0)=
\exp\{a{\rm ad}( F(Z))\}E^{AB}=E^{AB},
\eeq
and the transformation (\ref{trzZ}) is really anticanonical.
The inverse to this statement
is valid as well: an arbitrary set of anticanonical variables ${\bar Z}^A(Z)$
can be presented in the form
\beq
\label{genF}
{\bar Z}^A(Z)=\exp\{{\rm ad}( F(Z))\}Z^A
\eeq
with some generator functional $F(Z),\;\varepsilon(F(Z))=1 $. In Appendix B,
a proof of this fact is given.

Consider now a master-transformed action
$S'=S'(\varphi,\varphi^*)$ (\ref{trAc}). It was pointed out in
\cite{BLT-BV} that there are presentations of $S'$ in the following
forms
\beq
\label{Sprime1}
\exp\left\{\frac{i}{\hbar}S'\right\}=
\exp\{-[F,\Delta]\}\exp\left\{\frac{i}{\hbar}S\right\},
\eeq or
\beq
\label{Sprime2} S'=\exp\{{\rm ad}(
F)\}S+ i\hbar f({\rm ad}(F))\Delta F,
\eeq
where $S=S(\varphi,\varphi^{*})$, and $F=F(\varphi,\varphi^*)$
is a generator functional of an anticanonical
transformation, $f(x)=(\exp(x)-1)x^{-1}$. In accordance with
(\ref{trAc}), the first term in the right-hand side in
(\ref{Sprime2}) describes an anticanonical transformation of $S$ with an
odd generator functional $F$, while the second term is a half of a
logarithm of the Jacobian (\ref{Jacob}) of that transformation, up to
$(-i\hbar)$. In Appendix D, we give a proof of the latter statement.

Now we are in a position to study the properties of generating
functionals of Green functions
subjected to an arbitrary anticanonical transformation. We start with the
generating functionals of Green and connected Green functions
\beq
\label{ZWprime}
 {\cal Z}'={\cal Z}'(J,\varphi^{\ast})=\int d\varphi \exp\left\{\frac{i}{\hbar
}(S^{\prime}(\varphi,\varphi^{\ast})+J\varphi)\right\}=
\exp\left\{\frac{i}{\hbar}W^{\prime
}(J,\varphi^{\ast})\right\},
\eeq
where $S'$ is defined in (\ref{Sprime1}). The constructed generating functionals
(\ref{ZWprime}) obey the very important property of independence of $F$ for physical
quantities on-shell.\footnote{Note that in gauge theories the  "on-shell"
includes a definition of the physical state space.} Indeed, for infinitesimal
$\delta F$ the variation of ${\cal Z}'$
(\ref{ZWprime}),
\beq
\nonumber
&&\delta {\cal Z}'=-\frac{i}{\hbar}\int d\varphi
[(S,\delta F)-i\hbar(\Delta\delta F)]
\exp\left\{\frac{i}{\hbar
}(S(\varphi,\varphi^{\ast})+J\varphi)\right\}=\\
\label{infZprime}
&&=\frac{i}{\hbar}J_{A}\partial_{\varphi_{A}^{\ast}}\left[
\delta{\tilde F}(J,\varphi^{\ast})
\exp\left\{\frac{i}{\hbar}W(J,\varphi^{\ast})\right\}\right],
\eeq
is proportional to the external sources $J$. Due to the equivalence theorem
\cite{KT}, it means that the Green functions calculated with the help of
the generating functionals ${\cal Z}(J,\varphi^*)$ and ${\cal Z}'(J,\varphi^*)$
give the same physical answers on-shell.
 In deriving (\ref{infZprime}), the result
of calculation (\ref{delZ}) is used and the notation
\beq
\delta
{\tilde F}(J,\varphi^{\ast})=Z^{-1}(J,\varphi^*)\int d\varphi\;
\delta F(\varphi,\varphi^*) \exp\left\{\frac{i}{\hbar
}(S(\varphi,\varphi^{\ast})+J\varphi)\right\}
\eeq
 is introduced.

In the case of finite anticanonical transformations, we consider the
following anticanonically generalized action $S''$
\beq
\label{Sprime3}
\exp\left\{\frac{i}{\hbar}S''(\varphi,\varphi^{\ast})\right\}=
\exp\{-[F(\varphi ,\varphi^{\ast})+\delta
F(\varphi,\varphi^{\ast}),\Delta]\}
\exp\left\{\frac{i}{\hbar}S(\varphi,\varphi^{\ast})\right\},
\eeq
where $\delta F=\delta F(\varphi,\varphi^*)$ is an infinitesimal
functional. The  following representation holds:
\beq
\label{Fd}
\exp\{-[F(\varphi ,\varphi^{\ast})+\delta
F(\varphi,\varphi^{\ast}),\Delta]\}= \exp\{-[\delta{\cal
F}(\varphi,\varphi^{\ast}),\Delta]\}
\exp\{-[F(\varphi,\varphi^{\ast}),\Delta]\},
\eeq
where $\delta{\cal F}(\varphi,\varphi^{\ast})$ is defined by the relation
\beq
\label{dFc} \exp\{-{\rm ad}( F(\varphi,\varphi^{\ast}))-{\rm ad}
(\delta F(\varphi,\varphi^{\ast}))\} \exp\{-{\rm ad}
(F(\varphi,\varphi^{\ast}))\}= \exp\{-{\rm ad}(\delta {\cal
F}(\varphi,\varphi^{\ast}))\}.
\eeq
In Appendix C, a proof of
Eqs. (\ref{Fd}) and (\ref{dFc})  is given. Due to ({\ref{Fd}}), we
can present the action $S''$ in the form
\beq \label{S2S1}
\exp\left\{\frac{i}{\hbar}S''(\varphi,\varphi^{\ast})\right\}=
\exp\{-[\delta{\cal F}(\varphi,\varphi^{\ast}),\Delta]\}
\exp\left\{\frac{i}{\hbar}S'(\varphi,\varphi^{\ast})\right\}.
\eeq
Although we need here the infinitesimal functional
$\delta{\cal F}(\varphi,\varphi^{\ast})$,
the representation (\ref{S2S1}) by itself is valid
for arbitrary functional $\delta{\cal F}$.
In turn, the representation (\ref{S2S1}) allows us  the use of the previous
arguments concerning the case of infinitesimal anticanonical
transformations and for the statement that the generating functionals ${\cal Z}''$
and ${\cal Z}'$ constructed with the help of the actions $S''$ and $S'$,
respectively,  give the same physical results.

The next point of our study is connected with the behavior
of generating functionals
subjected to an arbitrary anticanonical transformation.
Consider a one-parameter family of functionals ${\cal Z}'(J,\varphi^{\ast};a)$,
\beq
&& \!\!\!\! {\cal Z}'(a)={\cal Z}^{\prime}(J,\varphi^{\ast};a)=\!
\int d\varphi \exp\left\{\frac{i}
{\hbar}(S'(\varphi,\varphi^{\ast};a)+J\varphi)\right\}=
\exp\left\{\frac{i}{\hbar}W^{\prime}(J,\varphi^{\ast};a)\right\}\!\!,\\
\label{Sprimea3}
&& \exp\left\{\frac{i}{\hbar}S'(\varphi,\varphi^{\ast};a)\right\}=
\exp\{-a[F(\varphi,\varphi^{\ast}),\Delta]\}
\exp\left\{\frac{i}{\hbar}S(\varphi,\varphi^{\ast})\right\},
\eeq
so that
\beq
{\cal Z}'(1)={\cal Z}'.
\eeq
Taking into account (\ref{delZ}) and (\ref{Sprimea3}), we derive the relation
\beq
\nonumber
&& \partial_{a}{\cal Z}'(a)=\frac{i}{\hbar}{\cal Z}'(a)\partial
_{a}W'(J,\varphi^{\ast};a)=\frac{i}{\hbar}{\cal Z}'(a)\partial
_{a}\Gamma(\varphi,\varphi^{\ast};a)=\\
\nonumber
&& =-\int d\varphi \exp\left\{\frac{i}{\hbar}J\varphi\right\}
[F(\varphi,\varphi^{\ast}),\Delta]
\exp\left\{\frac{i}{\hbar}S'(\varphi,\varphi^{\ast};a)\right\}=\\
\label{derZa}
&&=-\int d\varphi\exp\left\{\frac{i}{\hbar}J\varphi\right\}\Delta\left(F(\varphi
,\varphi^{\ast})\exp\left\{\frac{i}{\hbar}S'(\varphi,\varphi^{\ast}
;a)\right\}\right).
\eeq
By repeating similar calculations which lead us from (\ref{delZ})
to (\ref{anticGamma}) due to Eqs. (\ref{usID}), (\ref{usID1})
and (\ref{usR1})-(\ref{usRs}), we obtain
\beq
&&\partial_{a}\Gamma(\varphi,\varphi^{\ast};a)=(\mathcal{F}(\varphi
,\varphi^{\ast};a),\Gamma(\varphi,\varphi^{\ast};a)),\label{derfGam}\\
\label{calF}
&&\mathcal{F}(\varphi,\varphi^{\ast};a) =\left.  \frac{1}{{\cal Z}'
(J,\varphi^{\ast};a)}\int d{\tilde\varphi}
F({\tilde \varphi},\varphi^{\ast})\exp\left\{\frac{i}{\hbar
}(S'({\tilde \varphi},\varphi^{\ast};a)+
J{\tilde\varphi})\right\}\right|_{J=J(\varphi,\varphi^{\ast};a)}.
\eeq
We will refer to (\ref{derfGam}) as the basic equation describing dependence
of effective action
on an anticanonical transformation in the field-antifield formalism.
In Sect. 5, we present a
solution to this equation.
\\

\section{Solution to the basic equation}

In what follows below, we will use a short notation for all quantities depending
on the variables $\varphi,\varphi^*$,
\beq
\label{Eq.5.1}
\Gamma(\varphi,\varphi^*;a)\equiv \Gamma(a),\quad
\Gamma(\varphi,\varphi^*)\equiv \Gamma,\quad
{\cal F}(\varphi,\varphi^*;a)\equiv {\cal F}(a)
\eeq
and so on. Then the basic equation (\ref{derfGam}) is written as
\beq
\label{Eq.5.2}
\pa_a \Gamma(a)=({\cal F}(a),\Gamma(a))
={\rm ad}({\cal F}(a))\Gamma(a).
\eeq

We will study solutions to (\ref{Eq.5.2}) in the class
of regular functionals in $a$, by using
a power series expansion in this parameter. In the beginning,
let us find a solution to this equation
to the first order in $a$, presenting $\Gamma(a)$ and ${\cal F}(a)$ in the form
\beq
&&  \Gamma_{1}(a)\equiv\Gamma(a)=\Gamma+a\Gamma_{1|1}+O(a^{2}),\\
&&  \mathcal{F}_{1}(a)\equiv\mathcal{F}(a)=\frac{1}{a}\mathcal{F}_{1|1}(a)+O(a),
\;\mathcal{F}_{1|1}(a)=a\mathcal{F}_{1|1}.
\eeq
A straightforward calculation yields the following result
\beq
\Gamma_{1|1}=(\mathcal{F}_{1|1},\Gamma)={\rm ad}({\cal F}_{1|1})\Gamma.
\eeq
Introduce the notation $U_{1}(a)=\mathcal{F}_{1|1}(a)=
a\mathcal{F}_{1|1}$ and the functional $\Gamma_2(a)$ by the rule
\beq
\label{G2}
\Gamma_{2}(a)=\exp\{-{\rm ad}({U}_{1}(a))\}\Gamma_{1}(a).
\eeq
The dependence of $\Gamma_2(a)$ on $a$ is described by the equation
\beq
\label{EqG2}
\partial_{a}\Gamma_{2}(a)=(\mathcal{F}_{2}
(a),\Gamma_{2}(a))
\eeq
where
\beq
\label{F2}
 \mathcal{F}_{2}(a)=\left[\exp\{-a{\rm ad}(\mathcal{F}_{1|1})\}
 \mathcal{F}_{1}(a)-\mathcal{F}_{1|1}\right].
\eeq
It follows from (\ref{G2}) that the
functional $\Gamma_2(a)$ coincides with $\Gamma$ up to the second
order in $a$,
\beq
\Gamma_{2}(a)=\Gamma+O(a^{2})=\Gamma+a^{2}\Gamma_{2|2}+O(a^{3}).
\eeq
In turn, the functional ${\cal F}_2(a)$ vanishes to the first
order in $a$
\beq
\mathcal{F}_{2}(a)=O(a)=\frac{2}{a}\mathcal{F}_{2|2}(a)+O(a^{2}),
\;\mathcal{F}_{2|2}(a)=a^{2}\mathcal{F}_{2|2}.
\eeq
To the second
order in $a$, the solution to (\ref{EqG2}) reads
\beq
\Gamma_{2|2}=(\mathcal{F}_{2|2},\Gamma).
\eeq
Then, the functional
${\tilde \Gamma}_3(a)$ constructed by the rule
\beq
\tilde{\Gamma}_{3}(a)=\exp\{-{\rm ad}(\mathcal{F}
_{2|2}(a))\}\Gamma_{2}(a)
\eeq
coincides with $\Gamma$ up to the third order in $a$
\beq
\tilde{\Gamma}_{3}(a)=\Gamma+O(a^{3}).
\eeq
Introduce  the functional $\Gamma_3(a)$
\beq
\Gamma_{3}(a)=\exp\{-{\rm ad}(U_{2}(a))\}\Gamma_1(a),\quad
U_{2}(a)=\mathcal{F}_{1|1}(a)+\mathcal{F}_{2|2}(a).
\eeq
Note that
$\Gamma_3(a)$ coincides with ${\tilde \Gamma}_3(a)$ up to the third
order in $a$
\beq \label{G3ap}
\Gamma_{3}(a)=\tilde{\Gamma}_{3}(a)+O(a^{3})=\Gamma+O(a^{3})=
\Gamma+a^{3}\Gamma_{3|3}+O(a^{4}),
\eeq
so that we have
\beq
\nonumber
&&\Gamma_{3}(a)=\exp\{-{\rm ad}(\mathcal{F}_{2|2}(a))\}
\exp\{-{\rm ad}(\mathcal{F}_{1|1}(a))\}\Gamma_{1}(a)+O(a^3)=\\
&&=
\exp\{-{\rm ad}(\mathcal{F}_{2|2}(a))\} \Gamma_{2}(a)+O(a^3)
\eeq
due to the relation (\ref{Fn}). It follows from (\ref{G3ap}) that
\beq
\partial_{a}\Gamma_{3}(a)=3a^{2}\Gamma_{3|3}+O(a^{3}).
\eeq
On the other hand, we have
\beq
\label{eqG3}
\partial_{a}\Gamma_{3}(a)=(\mathcal{F}_{3}(a),\Gamma_{3}(a))=
{\rm ad}({\cal F}_3(a))\Gamma_3(a),
\eeq
where
\beq
\nonumber
\label{F3}
&&{\rm ad}(\mathcal{F}_{3}(a))=-\exp\{-{\rm ad}(U_{2}(a))\}
\partial_{a}\exp\{{\rm ad}(U_{2}(a))\}+\\
\label{F3}
&&-\exp\{-{\rm ad}(U_{2}(a))\}
{\rm ad}(\mathcal{F}_{1}(a))
\exp\{{\rm ad}(U_2(a))\},
\eeq
the operators on the right-hand side of (\ref{F3}) have certainly the form of
the ones of ${\rm ad}$, see Eqs. (\ref{X}), (\ref{Xt}) and (\ref{ABA}),
(\ref{eAB}). By using (\ref{AnA}), (\ref{Xza}), we derive from (\ref{F3})
and (\ref{eqG3})
\beq
\nonumber
&&{\rm ad}(\mathcal{F}_3(a))=-\frac{2}{a}{\rm ad}(\mathcal{F}_{2|2}(a))+
\exp\{-{\rm ad}(\mathcal{F}_{2|2}(a))\}{\rm ad}(\mathcal{F}_{2}(a))
\exp\{{\rm ad}(\mathcal{F}_{2|2}(a))\}+O(a^{2})=\\
&& =\frac{3}{a}{\rm ad}(\mathcal{F}_{3|3}(\varphi,\varphi^{\ast};a))+O(a^{3}%
),\qquad{\rm ad}(\mathcal{F}_{3|3}(a))=a^{3}{\rm ad}(\mathcal{F}_{3|3}),\\
 &&\qquad\qquad\qquad\qquad\qquad \Gamma_{3|3}=(\mathcal{F}_{3|3},\Gamma).
\eeq

Suppose that on the {\it n}th step of our procedure we
have obtained the following relations,
\beq
\nonumber
&&  \Gamma_{n}(a)=\exp\{-{\rm ad}(U_{n-1}(a))\}\Gamma_{1}(a)=\Gamma+O(a^{n})=\\
&& =\Gamma+a^{n}\Gamma_{n|n}+O(a^{n+1}),\;
U_{n-1}(a)=\sum_{k=1}^{n-1}\mathcal{F}_{k|k}(a)\equiv\mathcal{F}_{[n-1|n-1]}(a),\\
&& \qquad\qquad\qquad\partial_{a}\Gamma_{n}(a)=(\mathcal{F}_{n}
(a),\Gamma_{n}(a)),\\
&&  \mathcal{F}_{n}(a)=O(a^{n})=\frac{n}{a}
\mathcal{F}_{n|n}(a)+O(a^{n+1}),\quad \mathcal{F}_{n|n}(a)=
a^{n}\mathcal{F}_{n|n},\\
&& \qquad\qquad\qquad\qquad \Gamma_{n|n}=(\mathcal{F}_{n|n},\Gamma).
\eeq
We set
\beq
U_{n}(a)=\mathcal{F}_{[n|n]}(a).
\eeq
Then we have
\beq
&& \exp\{-{\rm ad}(\mathcal{F}_{n|n}(a)\}\Gamma_{n}(a)=\Gamma+O(a^{n+1}),\\
\nonumber
&&  \Gamma_{n+1}(a)=\exp\{-{\rm ad}(U_{n}(a))\}\Gamma_{1}(a)=
\exp\{-{\rm ad}(\mathcal{F}_{n|n}(a))\}\Gamma_{n}(a) +O(a^{n+1})=\\
&&=\Gamma+O(a^{n+1})=\Gamma+a^{n+1}\Gamma_{n+1|n+1}+O(a^{n+2}).
\eeq
In  a similar manner, we derive the equation for $\Gamma_{n+1}(a)$
\beq
\partial_{a}\Gamma_{n+1}(a)=(\mathcal{F}_{n+1}(a),\Gamma_{n+1}(a)),
\eeq
where
\beq
\nonumber
&&{\rm ad}(\mathcal{F}_{n+1}(a))=-\exp\{-{\rm ad}(U_{n}(a))\}
\partial_{a}\exp\{{\rm ad}(U_{n}(a))\}+\\
&&+
\exp\{-{\rm ad}(U_{n}(a))\}{\rm ad}(\mathcal{F}_{1}(a))\exp\{{\rm ad}(U_{n}(a))\}.
\eeq
By the same reasons used at the previous stages, we conclude that
\beq
\nonumber
&&{\rm ad}(\mathcal{F}_{n+1}(a))=-\frac{n}{a}{\rm ad}(\mathcal{F}_{n|n}(a))
+\exp\{-{\rm ad}(\mathcal{F}_{n|n}(a))\}{\rm ad}(\mathcal{F}_{n}(a))
\exp\{{\rm ad}(\mathcal{F}_{n|n}(a))\}+O(a^{n})=\\
&&=\frac{n+1}{a}{\rm ad}(\mathcal{F}_{n+1|n+1}(a))+O(a^{n+1}),\quad
\mathcal{F}_{n+1|n+1}(a)=a^{n+1}\mathcal{F}_{n+1|n+1},\\
&&\qquad\qquad\qquad\qquad  \Gamma_{n+1|n+1}=(\mathcal{F}_{n+1|n+1},\Gamma),\\
&&
\Gamma(a)=\exp\{{\rm ad}(U_{n}(a))\}\Gamma_{n+1}(a)=\exp\{{\rm ad}({U}_{n}(a))\}
\Gamma+O(a^{n+1}).
\eeq
Finally, by applying the induction method,
we obtain that a solution to the basic equation (\ref{Eq.5.2}) can
be presented in the form
\beq
\label{canGf}
\Gamma(a)=\exp\{{\rm ad}({U}(a))\}\Gamma,
\eeq
which is nothing but an
anticanonical transformation of $\Gamma$ with a generator functional
$U(a)$ defined by functional $\mathcal{F}(a)$ in (\ref{Eq.5.2}) as
\beq
U(a)=\sum_{k=1}^{\infty}\mathcal{F}_{k|k}(a).
\eeq
In this proof, we have found a possibility to express the
relation between $U(a)$ and $\mathcal{F}(a)$ in the form
\beq
\label{calF1}
\mathcal{F}(a)=-\exp\{{\rm ad}({U}(a))\}\partial_{a}\exp\{-{\rm ad}({U}(a))\}.
\eeq
In turn, the relation (\ref{calF1}) can be considered as a new
representation of the functional (\ref{calF}). Let us notice that
the functional $U(a)$ in (\ref{canGf}) depends on the functional
$F(a)$ only and does not depend on the choice of an initial data
for $\Gamma(a)$. The above formulae (\ref{canGf}) - (\ref{calF1})
just represent the important relationship between the ordinary
exponential and the path-ordered one.

Let us state again that the dependence of the effective action on a
finite anticanonical transformation
with a generating functional $Y(\varphi,\Phi^*;a)$ is really described in terms
of anticanonical transformation with a generator functional
$U(\varphi,\varphi^*;a)$. As an anticanonical transformation is a change
of variables in $\Gamma$,
in particular, it means that, on-shell, the effective action does not depend
on gauges introducing with the help of anticanonical transformations.
\\

\section{Discussions}

In the present article, we have explored  a conception of a gauge
fixing procedure in the field-antifield formalism \cite{BV,BV1},
based on the use of anticanonical transformations of general
type. The approach includes an action (master-transformed
action) constructed with the help of anticanonical master transformation
and being non-degenerate. The master-transformed action is a sum of two terms:
one is an action subjected
to an anticanonical transformation and the other is a term
connecting with a logarithm of a superdeterminant of this
anticanonical transformation. This action satisfies the quantum
master equation \cite{BLT4,BBD} (see also Appendix D). The generating
functionals of the Green functions constructed via the
master-transformed action obey the important property
of the gauge independence of physical quantities on-shell, and they
satisfy the Ward identity. We have found that any (finite)
anticanonical master transformation of an action leads to the corresponding
anticanonical transformation of effective action (generating
functional of vertex functions) provided the generating
functional of Green functions is constructed with the help of an
anticanonical master action. We have proved the existence of
a generator functional of an anticanonical transformation of the
effective action. This result is essential when proving the
independence of the effective action of anticanonical
transformations on-shell and, on the other hand, it may supplement
in a non-trivial manner the representation of anticanonical
transformations in the form of a path-ordered exponential
\cite{BBD}.
\\

\section*{Acknowledgments}
\noindent
 I. A. Batalin would like  to thank Klaus Bering of Masaryk
University for interesting discussions. The work of I. A. Batalin is
supported in part by the RFBR grants 14-01-00489 and 14-02-01171.
 The work of P. M. Lavrov is supported by the Ministry of Education and Science of
Russian Federation, project No Z.867.2014/K.  The work of
I. V. Tyutin is partially supported by the RFBR grant 14-02-01171.
\\

\appendix
\section*{Appendix A: Infinitesimal variation of effective action}
%\section{}
\setcounter{section}{1}
\renewcommand{\theequation}{\thesection.\arabic{equation}}

Here we prove the possibility to present the equation (\ref{delGamma})
in the form (\ref{anticGamma}). To do this,
 we introduce the matrix of second derivatives of $\Gamma$, $\Gamma_{ij}$,
and its inverse, $M^{ij}$,
\beq
\label{usR1}
&&\Gamma_{ij}    \equiv\partial_{\varphi^{i}}\partial_{\varphi^{j}}
\Gamma=(-1)^{\varepsilon_i\varepsilon_j}\Gamma_{ji},
\quad \varepsilon(\Gamma_{ij})= \varepsilon
_{i}+\varepsilon_{j},\\
\label{usR2}
&&M^{ij}\Gamma_{jk} =\delta_{k}^{i},\quad \varepsilon(M^{ij})= \varepsilon
_{i}+\varepsilon_{j},\quad M^{ji}=
(-1)^{\varepsilon_i\varepsilon_j+\varepsilon_i+\varepsilon_j}M^{ij}.
\eeq
From the Ward identity (\ref{WIG}) written in the form
\beq
\Gamma_i\Gamma^{i^*}=0,\quad \Gamma^{i^*}=\Gamma\overleftarrow{\pa}_{\varphi^*_i},
\quad \Gamma_i=\pa_{\varphi^i}\Gamma ,
\eeq
it follows that the relations
\beq
(-1)^{\varepsilon_j\varepsilon_k+\varepsilon_k}\Gamma_{k}\Gamma_{j}^{k^{\ast}}=
(-1)^{\varepsilon_j}\Gamma^{k^{\ast}}
\Gamma_{kj},\quad \Gamma_{j}^{k^{\ast}}=\partial_{\varphi^{j}}\partial
_{\varphi_{k}^{\ast}}\Gamma
\eeq
hold. By taking these relations into account, we have
\beq
\nonumber
&&\,(-1)^{\varepsilon_k}\Gamma_{k}\left[  \partial_{\varphi_{k}^{\ast}}J_{j}
(\varphi,\varphi^{\ast})\right]  =-(-1)^{\varepsilon_j+\varepsilon_k}
\Gamma_{k}\partial_{\varphi
_{k}^{\ast}}\partial_{\varphi^{j}}\Gamma=\\
 &&=-(-1)^{\varepsilon_j\varepsilon_k+\varepsilon_k}\Gamma_{k}\Gamma_{j}^{k^{\ast}}
 =-(-1)^{\varepsilon_j}\Gamma^{k^{\ast}}\Gamma_{kj}
\eeq
and
\beq
\nonumber
&&\partial_{\varphi^{k}}\mathcal{X}=\left[  \partial_{\varphi^{k}}%
J_{j}(\varphi,\varphi^{\ast})\right]  \partial_{J_{j}}\tilde{X}=
-(-1)^{\varepsilon_j}
\Gamma_{kj}\partial_{J_{j}}\tilde{X}\;\Longrightarrow\\
&&  \left.  \partial_{J_{j}}\tilde{X}(J,\varphi^{\ast})\right|  _{J=J(\varphi
,\varphi^{\ast})}=-(-1)^{\varepsilon_j}M^{jk}\partial_{\varphi^{k}}\mathcal{X}
(\varphi,\varphi^{\ast}).
\eeq
Therefore
\beq
\nonumber
&&  \,(-1)^{\varepsilon_k}\Gamma_{k}\left[  \partial_{\varphi_{k}^{\ast}}J_{j}
(\varphi,\varphi^{\ast})\right]  \left.  \partial_{J_{j}}\tilde{X}
(J,\varphi^{\ast})\right|  _{J=J(\varphi,\varphi^{\ast})}=\\
\label{usRs}
&&=\Gamma^{k^{\ast}}\Gamma_{kj}M^{jk}\partial_{\varphi^{k}}\mathcal{X}%
(\varphi,\varphi^{\ast})=\Gamma\overleftarrow{\partial}_{\varphi_{k}^{\ast}}
\partial_{\varphi^{k}}\mathcal{X}.
\eeq
Substituting (\ref{usRs}) in (\ref{delGamma}), we have
derived the  equation (\ref{anticGamma}) for a variation
of $\Gamma$.
\\

\appendix
\section*{Appendix B: Generator of anticanonical transformation}
%\section{}
\setcounter{section}{2}
\renewcommand{\theequation}{\thesection.\arabic{equation}}
\setcounter{equation}{0}

Here we give a proof that any anticanonical transformation
%with a generating functional $Y$
can be described by the corresponding generator ${\rm ad}( F)$
in the sense of (\ref{genF}). Firstly,
we note that if  ${\bar Z}_{l}^{A}(Z)$, $l=1,2$, are anticanonical variables,
\beq
({\bar Z}_{1}^{A}(Z),{\bar Z}_{1}^{B}(Z))=
({\bar Z}_{2}^{A}(Z),{\bar Z}_{2}^{B}(Z))=E^{AB}\;,
\eeq
then the compositions of these variables, ${\bar Z}_{12}^{A}(Z)=
{\bar Z}_{1}^{A}({\bar Z}_{2}(Z))$ and ${\bar Z}_{21}^{A}
(Z)={\bar Z}_{2}^{A}({\bar Z}_{1}(Z))$, are anticanonical as well. Indeed, we have
\beq
\nonumber
&& ({\bar Z}_{12}^{A}(Z),{\bar Z}_{12}^{B}(Z))=
{\bar Z}_{12}^{A}(Z)\overleftarrow{\partial}_{C}E^{CD}
\overrightarrow{\partial}_{D}{\bar Z}_{12}^{B}(Z)=\\
\nonumber
&&={\bar Z}_{1}^{A}({\bar Z}_{2})\overleftarrow{D}_{2|M}
\left[  {\bar Z}_{2}^{M}(Z)\overleftarrow
{\partial}_{C}E^{CD}\overrightarrow{\partial}_{D}{\bar Z}_{2}^{N}(Z)\right]
\overrightarrow{D}_{2|N}{\bar Z}_{1}^{B}({\bar Z}_{2})=\\
&&={\bar Z}_{1}^{Z}({\bar Z}_{2})\overleftarrow{D}_{2|M}E^{MN
}\overrightarrow{D}_{2|N}
{\bar Z}_{1}^{B}({\bar Z}_{2})=E^{AB},\;\;D_{2|A}=
\frac{\partial}{\partial {\bar Z}_{2}^{A}}\;.
\eeq
In particular, the
variables ${\bar Z}_{12}^{A}(Z)=\exp\{{\rm ad}F(Z)\}{\bar Z}_{1}^{A}(Z)$
are anticanonical if ${\bar Z}_{1}^{A}(A)$ are anticanonical variables.
 Indeed, we have
\beq
{\bar Z}_{12}^{Z}(Z)={\bar Z}_{1}^{Z}({\bar Z}_{2}(Z)),
\quad {\bar Z}_{2}^{A}(Z)=\exp\{{\rm ad}F(Z)\}Z^{A}.
\eeq
Secondly, the next remark is obvious
\beq
\label{Fn}
&&\exp\{{\rm ad}(F_{[n]}(Z;a))\}\exp\{{\rm ad}(F_{n+1}(Z;a))\}=
\exp\{{\rm ad}(F_{[n+1]}(Z;a))\}+O(a^{n+2}),\\
\nonumber &&F_{[k]}(Z;a)=\sum_{l=1}^kF_l(Z;a),\quad
F_l(Z;a)=a^{l}F_{l}(Z). \eeq

Now let ${\bar Z}^A(Z;a)\equiv {\bar Z}_{1}^{A}(Z;a)=Z^{A}+
a Z_{1|1}^{A}(Z)+O(a^{2})$ be
anticanonical variables with a generating functional $Y(\varphi,\Phi^{\ast};a)
\equiv Y_{1}(\varphi,\Phi^{\ast};a)=\Phi_{i}^{\ast}\varphi^{i}-
a f_{1|1}(\varphi,\Phi^{\ast})+O(a^{2})$.
Taking into account (\ref{Ya}) - (\ref{Phi*a}) and (\ref{Fsc2}) - (\ref{Fhat}),
we have
\beq
&&Z_{1|1}^{A}(Z) =(F_{1|1}(Z),Z^{A}),\;F_{1|1}(Z)=f_{1|1}(\varphi
,\varphi^{\ast})\;\Longrightarrow\\
&&{\bar Z}_{1}^{A}(Z;a) =\exp\{{\rm ad}(F_{1|1}(Z;a))\}Z^{A}+O(a^{2}).
\eeq
Then we introduce (anticanonical) variables ${\bar Z}_{2}^{A}(Z;a)$,
\beq
{\bar Z}_{2}^{A}(Z;a)=\exp\{-{\rm ad}(F_{1|1}(Z;a))\}{\bar Z}_{1}^{A}(Z;a)=Z^{A}
+a^{2}Z_{2|2}^{A}(Z)+O(a^{3}),
\eeq
with the corresponding generating functional
\beq
Y_{2}(\varphi,\Phi^{\ast};a)=
\Phi_{i}^{\ast}\varphi^{i}-a^{2}f_{2|2}(\varphi,\Phi^{\ast})+O(a^{3}).
\eeq
As a result, we have
\beq
&& Z_{2|2}^{A}(Z)=(F_{2|2}(Z),Z^{A}),\;F_{2|2}(Z)=f_{2|2}(\varphi,\varphi
^{\ast})\;\Longrightarrow\\
&& {\bar Z}_{2}^{A}(Z;a)=\exp\{{\rm ad}({F}_{2|2}(Z;a))\}Z^{A}+
O(a^{3})\;\Longrightarrow\\
\nonumber
&& {\bar Z}_{1}^{A}(Z;a)=\exp\{{\rm ad}({F}_{1|1}(Z;a))\}{\bar Z}_{2}^{A}(Z;a)=\\
\nonumber
&&=\exp\{{\rm ad}({F}_{1|1}(Z;a))\}\exp\{{\rm ad}({F}_{2|2}(Z;a))\}Z^{A}+O(a^{3})=\\
&& =\exp\{{\rm ad}({F}_{[2|2]}(Z;a))\}Z^{A}+O(a^{3}),
\eeq
where the relation (\ref{Fn}) is used.

Suppose that a representation of anticanonical variables ${\bar Z}_{1}^{A}(Z;a)$
does exist in the form
\beq
{\bar Z}_{1}^{A}(Z;a)=\exp\{{\rm ad}({F}_{[n|n]}(Z;a))\}Z^{A}+O(a^{n+1}),\;
{\rm ad}({F}_{[n|n]}(Z;a))=\sum_{k=1}^{n}{\rm ad}({F}_{k|k}(Z;a)).
\eeq
Introduce the (anticanonical) variables ${\bar Z}_{n+1}^{A}(Z;a)$,
\beq
{\bar Z}_{n+1}^{A}(Z;a)=\exp\{-{\rm ad}({F}_{[n|n]}(Z;a))\}{\bar Z}_{1}^{A}(Z;a)=
Z^{A}+a^{n+1}Z_{n+1|n+1}^{A}(Z)+O(a^{n+2}).
\eeq
The corresponding generating functional $Y_{n+1}(\varphi,\Phi^{\ast};a)$
has the form
\beq
Y_{n+1}(\varphi,\Phi^{\ast};a)=\Phi_{i}^{\ast}\varphi
^{i}-a^{n+1}f_{n+1|n+1}(\varphi,\Phi^{\ast})+O(a^{n+2}).
\eeq
By the usual manipulations, we find
\beq
&& Z_{n+1|n+1}^{A}(Z)=(F_{n+1|n+1}(Z),Z^{A}),\;F_{n+1|n+1}(Z)=f_{n+1|n+1}
(\varphi,\varphi^{\ast}),\\
&& {\bar Z}_{n+1}^{A}(Z;a)=\exp\{{\rm ad}({F}_{n+1|n+1}(Z;a))\}Z^{A}
+O(a^{n+2}),\\
\nonumber
&& {\bar Z}_{1}^{A}(Z;a)=\exp\{{\rm ad}({F}_{[n|n]}(Z;a))\}{\bar Z}_{n+1}^{A}(Z;a)=\\
\nonumber
&&=\exp\{{\rm ad}({F}_{[n|n]}(Z,a))\}\exp\{{\rm ad}({F}_{n+1|n+1}(Z;a))\}
Z^{A}+O(a^{n+2})=\\
&&=\exp\{{\rm ad}({F}_{[n+1|n+1]}(Z;a))\}Z^{A}+O(a^{n+2}).
\eeq
Applying the induction method, we have proved
that an arbitrary set of anticanonical
variables ${\bar Z}^{A}(Z)$ can be really represented in the form (\ref{genF}).
\\

\appendix
\section*{Appendix C: Some useful formulas}
%\section{}
\setcounter{section}{3}
\renewcommand{\theequation}{\thesection.\arabic{equation}}
\setcounter{equation}{0}

Consider a set of differential operators
${\rm ad}( A(Z)), {\rm ad}(B(Z)),...$,
$\varepsilon(A(Z))=1, \varepsilon(B(Z))=1,...$ applied
to any functional $M(Z)$ of anticanonical variables
$Z=(\varphi,\varphi^*)$ as the left adjoint of the antibracket.
If a multiplication operation is introduced as the commutator, then this
set can be considered as a Lie superalgebra. Indeed, due to the
symmetry properties and the Jacobi identity for the antibracket, we
have
\beq
\label{comAB}
&&[{\rm ad}( A(Z)),{\rm ad}( B(Z)]={\rm ad}({A}(Z)){\rm ad}({B}(Z))
-{\rm ad}({B}(Z)){\rm ad}({A}(Z))={\rm ad}({C}_{A|B}(Z)),\\
&&C_{A|B}(Z)=(A(Z),B(Z)),\quad \varepsilon(C_{A|B}(Z)=1,
\eeq
or, in  more detail, by application to $M(Z)$,
\beq
\nonumber
&&(A(Z),(B(Z),M(Z)))-(B(Z),(A(Z),M(Z)))=\\
\nonumber
&&(A(Z),(B(Z),M(Z)))+(B(Z),(M(Z),A(Z)))=\\
\label{antAB}
&&=-(M(Z),(A(Z),B(z)))=((A(Z),B(Z)),M(Z))={\rm ad}({C}_{A|B}(Z))M(Z).
\eeq
Note that the operators under consideration give a good example of odd first-order
differential operations which are not nilpotent, $({\rm ad}( A(Z)))^2\neq 0$.

It is obvious that
\beq
\label{AnA}
&& \exp\{{\rm ad}({A}_{n+1}(a))\}\exp\{{\rm ad}({A}_{[n]}(a))\}=
\exp\{{\rm ad}({A}_{[n+1]}(a))\}
+O(a^{n+2}),\\
&&  A_{[n]}(a)=\sum_{k=1}^{n}A_{k}(a),\quad A_{k}(a)=a^{k}A_{k}
\eeq
(see, also (\ref{Fn})).

Taking into account a series expansion
\beq
\nonumber
&&\exp\{{\rm ad}( A(Z))\}{\rm ad}( B(Z))\exp\{-{\rm ad}( A(Z))\}=
{\rm ad}( B(Z))+[{\rm ad}( A(Z)),{\rm ad}( B(Z))]+\\
&&+
\frac{1}{2!}[{\rm ad}( A(Z)),[{\rm ad}( A(Z)),{\rm ad}( B(Z))]]+\cdots,
\eeq
using relations similar to (\ref{comAB}) - (\ref{antAB}) and Jacobi
identity for the  antibracket, we deduce the identity
\beq
\label{ABA}
&& \exp\{{\rm ad}({A}(Z))\}{\rm ad}({B}(Z))\exp\{-{\rm ad}({A}(Z))\}=
{\rm ad}({D}_{A|B}(Z)),\\
\nonumber
&& D_{A|B}(Z)=B(Z)+(A(Z),B(Z))+\frac{1}{2!}(A(Z),(A(Z),B(Z)))+
\cdots=\\
\label{eAB}
&&=\exp\{{\rm ad}({A}(Z))\}B(Z),\quad \varepsilon(D_{A|B}(Z))=1 .
\eeq

The useful identity
\beq
&&
\label{Xza}
X=X(Z;a)=\exp\{{\rm ad}({A}(Z;a))\}
\partial_{a}\exp\{-{\rm ad}({A}(Z;a))\}=
-{\rm ad}({D}_{A}(Z;a)),\\
&& D_{A}(Z;a)=f({\rm ad}({A}(Z;a)))\partial_{a}A(Z;a),\\
&& f(x)=\big(\exp(x)-1\big)x^{-1},\quad
\varepsilon(A(Z;a))=1,\;\varepsilon(D_{A}(Z;a))=1,
\eeq
holds, as well. Indeed, let us introduce an operator $X(t)$,
\beq \label{X}
X(t)=X(Z;a;t)=\exp\{t{\rm ad}({A}(Z;a))\}\partial_{a}
\exp\{-t{\rm ad}({A}(Z;a))\},\; X(0)=0,\;X(1)=X.
\eeq
Then we have
\beq
\nonumber
&&\partial_{t}X(t)=-\exp\{t{\rm ad}({A}(Z;a))\}{\rm ad}(\partial_{a}A(Z;a))
\exp\{-t{\rm ad}({A}(Z;a))\}=\\
\label{Xtad}
&&=-{\rm ad}({C}_{\partial_{a}A}(Z;a;t)),\\
\label{Cta}
&&  C_{\partial_{a}A}(Z;a;t)=\exp\{t{\rm ad}({A}(Z;a))\}\partial_{a}A(Z;a).
%\;\Longrightarrow
\eeq
In deriving (\ref{Xtad}) and (\ref{Cta}), the identities (\ref{ABA}) and (\ref{eAB})
are used. Using initial data for $X(t)$, it follows from (\ref{Xtad}) that
\beq \label{Xt}
X(t)=-t{\rm ad}({D}_{\partial_{a}A}(Z;a;t)),\quad
D_{\partial_{a}A}(Z;a;t)=f(t{\rm ad}({A}(Z;a)))\partial_{a}A(Z;a).
\eeq

We will use the following convention and notation for applying operators
$R$ and $\hat{R}$,
\beq
F(R)A(Z)(...)=[F(R)A(Z)](...),\;F(\hat{R})A(Z)(...)=F(R)[A(Z)(...)],
\eeq
where $F(R)=F(x)|_{x=R}$ , $A(Z)$ is a function and $(...)$ means an arbitrary
quantity.

Consider a first-order differential operator
\beq
N(Z)\pa\equiv N^A(Z)\pa_A,\quad \pa_A=\frac{\pa}{\pa Z^A},\quad
\varepsilon(N^A(Z))=\varepsilon(Z^A),
\eeq
where $N^A(Z)$ are some functionals of $Z$. Let
\beq
\label{ZNz}
{\bar Z}^A(Z)\equiv \exp\{N(Z)\partial\}Z^A,
\eeq
then we have
\beq
\nonumber
&&\exp\left\{N(Z)\hat{\partial}\right\}Z^A\exp\left\{-N(Z)\hat{
\partial}\right\}=\sum_{k=0}\frac{1}{k!}[N(Z)\hat{\partial}
,[N(Z)\hat{\partial},...[N(Z)\hat{\partial},Z^A]...]]_{k\;\mathrm{times}}
= \\
&&=\sum_{k=0}\frac{1}{k!}[N(Z)\partial]^{k}Z^A=
\exp\{N(Z)\partial\}Z^A={\bar Z}^A(Z)
\eeq
where the relation
\beq
[N(Z)\hat{\partial},M(Z)]=N(Z)\partial M(Z)
\eeq
is used. In general
\beq
\label{NZz}
\exp\left\{N(Z)\hat{\partial}\right\}g(Z)
\exp\left\{-N(Z)\hat{\partial}\right\}
=\exp\{N(Z)\partial\}g(Z)=g({\bar Z}).
\eeq

Consider a more general differential operator than in (\ref{ZNz}),
\beq
\label{La}
L(a)=\exp\left\{aM(Z)+aN(Z)\hat{\partial}\right\}
\eeq
where $M(Z)$ is a functional of $Z$ and $a$ is a parameter.
We prove that there is
a representation of this operator in the form
\beq
\label{LHa}
L(a)=H(Z,a)\exp\left\{aN(Z)\hat{\partial}\right\}
\eeq
where $H(Z,a)$ is a functional. Indeed, it follows
from (\ref{La}) and (\ref{LHa}) that
\beq
H(Z,a)=\exp\left\{aM(Z)+aN(Z)\hat{\partial}\right\}
\exp\left\{-aN(Z)\hat{\partial}\right\}.
\eeq
By differentiating $H(Z,a)$ with respect to $a$, one gets the relation
\beq
\frac{d^{n}}{da^{n}}H(Z,a)=\exp\left\{aM(Z)+aN(Z)\hat{\partial}
\right\}h_{n}\exp\left\{-aN(Z)\hat{\partial}\right\},
\eeq
where
\beq
h_{n}=\left(M(Z)+N(Z)
\hat{\partial}\right)h_{n-1}-h_{n-1}N(Z)\hat{\partial}, \quad
h_{0}=1,\;h_{1}=M(Z).
\eeq
Suppose that $h_k, 0\leq k\leq n$ are some functionals, then
\beq
h_{n+1}=M(Z)h_{n}+N(Z)\hat{\partial}h_{n}-h_{n}N(Z)\hat{\partial}
=M(Z)h_{n}+N(Z)\partial h_{n}
\eeq
is a functional, as well. The latter means that all $a$-derivatives of $H(Z,a)$
taken at $a=0$ are some functionals too and, as a consequence, $H(Z,a)$
is a functional.

Now we can derive a representation for $H(Z,a)$. We start with the equation
\beq
\frac{d}{da}H(Z,a)=\exp\left\{aM(Z)+aN(Z)\hat{\partial}
\right\}M(Z)\exp\left\{-aN(Z)\hat{\partial}\right\},
\eeq
which can be rewritten as
\beq
\nonumber
\frac{d}{da}H(Z,a)&=&H(Z,a)\left(\exp\left\{aN(Z)\hat{\partial}
\right\}M(Z)\exp\left\{-aN(Z)\hat{\partial}\right\}\right)=\\
&=&\Big(\exp\left\{aN(Z)\partial\right\}M(Z)\Big)H(Z,a)
\eeq
where the relation (\ref{NZz}) is used. Integrating this equation leads to
\beq
H(Z)=H(Z,1)=\exp [f(x)M(Z)],\;f(x)=\frac{\exp (x)-1}{x},\;x=N(Z)\partial.
\eeq
Finally, we have
\beq
\exp\left\{M(Z)+N(Z)\hat{\partial}\right\}=
\exp [f(x)M(Z)]\exp\left\{N(Z)
\hat{\partial}\right\},\;x=N(Z)\partial .  \label{2.9}
\eeq

\appendix
\section*{Appendix D: Master-transformed actions}
%\section{}
\setcounter{section}{4}
\renewcommand{\theequation}{\thesection.\arabic{equation}}
\setcounter{equation}{0}

Here we present a set of properties concerning master-transformed actions.

Firstly, we prove that an action $S'$ constructed by the rule (\ref{Sprime1})
from $S$, being a
solution to the quantum master equation,
 satisfies the quantum master equation, as well. To do this, we
 consider a functional $X(Z)$ and the transformation $X(Z)\rightarrow
X^{\prime }(Z)=X(Z,1)$ of the form%
\begin{equation}
\exp \left\{\frac{i}{\hbar}X(Z,a)\right\}=\exp \big\{-a[F(Z),{\hat\Delta}
]\big\}\exp \left\{\frac{i}{\hbar}X(Z)\right\},\quad X(Z,0)=X(Z).
\label{3.3}
\end{equation}
The transformation (\ref{3.3}) has the property:
\beq
\Delta \exp \left\{\frac{i}{\hbar}X(Z)\right\}=0\;\Longrightarrow
\Delta \exp \left\{\frac{i}{\hbar}X(Z,a)\right\}=0.
\eeq
Indeed, let us introduce a functional
\beq
Y(Z,a)=\Delta\exp\left\{\frac{i}{\hbar}X(Z,a)\right\},\quad
Y(z,0)=\Delta\exp\left\{\frac{i}{\hbar}X(Z)\right\}.
\eeq
Then we have
\beq
\frac{d}{da}Y(Z,a)=-
{\hat\Delta}\left([F(Z),\Delta]\right)\exp\left\{\frac{i}{\hbar}X(Z,a)\right\}=
-{\hat\Delta} F(Z)Y(Z,a)
\eeq
where the nilpotency of $\Delta$ operator is used.
Integrating this equation gives
\beq
&&Y(Z,a)=
\exp\{-a{\hat\Delta} F(Z)\}Y(Z,0)\;
\Longrightarrow \\
&&\Delta\exp\left\{\frac{i}{\hbar}X(Z,a)\right\}=\exp\{-a{\hat\Delta} F(Z)\}
\Delta\exp\left\{\frac{i}{\hbar}X(Z)\right\}.
\eeq

Secondly, to prove the presentation of (\ref{Sprime2}), we consider the relation
(\ref{3.3}) in more detail. Note that
\beq
\label{FG}
[F(z),\Delta ]=(\Delta F(Z))-{\rm ad}({F}(Z)),
\eeq
and we have the following identification of (\ref{FG}) with the functions $M(Z)$
and the operator $N^{A}(Z)\partial _{A}$ from
 (\ref{LHa})
\beq
M(Z)=-\Delta F(Z),\;N^{A}(Z)\partial_A={\rm ad}({F}(Z)).
\eeq
It follows from (\ref{2.9}) that
\beq
X^{\prime}=\exp\{{\rm ad}({F}(Z))\}X+i\hbar f({\rm ad}({F}(Z)))\Delta F.
\label{3.9}
\eeq
In the right-hand  side in (\ref{3.9}), the first term is an anticanonical
transformation with finite fermionic generator $F$, while the second term is
a half of a logarithm of the Jacobian of that transformation, up to $(-i\hbar)$.
It is obvious that the inverse statement holds as well: the validity of the
relation (\ref{3.9}) implies Eq. (\ref{3.3}).

Now we show that  an equality holds of
\begin{equation} \label{3.10}
\exp\{-[F_2(Z)+F_1(Z),\Delta]\}=\exp\{-[{\cal F}_2(Z),\Delta]\}
\exp \{-[F_1(Z),\Delta]\},
\end{equation}
where ${\cal F}_2(Z)$ is determined by the relation
\begin{equation}
\label{F1-F3}
\exp\{[{\rm ad}({F}_2(Z))+{\rm ad}({F}_1(Z))]\}\exp\{-{\rm ad}({F}_1(Z))\}=
\exp\{{\rm ad}({{\cal F}}_2(Z))\}.
\end{equation}
The existence of Eqs. (\ref{3.10}) and (\ref{F1-F3})
means that transformations generated by
$\exp\{-[F(Z),\Delta]\}$ and $\exp\{{\rm ad}({F}(Z))\}$ obey a group property.

Consider anticanonical transformations generated by Fermionic functions $F_1(Z)$,
$F_1(Z)+F_2(Z)$ and ${\cal F}_2(Z)$
\beq
\label{trF1F2}
&&{\bar Z}^A_1(Z)=\exp\{{\rm ad}({F}_1(Z))\}Z^A,\;
{\bar Z}^A_2(Z)=\exp\{[{\rm ad}({F}_2(Z))+{\rm ad}({F}_1(Z))]\}Z^A,\\
\label{trF3}
&&{\bar{\cal Z}}^A_2(Z)=\exp\{{\rm ad}({\cal F}_2(Z))\}Z^A.
\eeq
Then, due to (\ref{F1-F3}), we have
\begin{eqnarray}
\label{Z122}
{\bar Z}^A_2(Z)=\exp\{{\rm ad}({{\cal F}}_2(Z))\}\exp\{{\rm ad}({F}_1(Z))\}Z^A=
{\bar Z}^A_1({\bar{\cal Z}}_2(Z)),
\end{eqnarray}
For a given action $S(Z)$, the relations
\begin{eqnarray}
&& S_1(Z)=\exp\{{\rm ad}({F}_1(Z))\}S(Z)=S({\bar Z}_1(Z)),\\
\nonumber
&&S_2(Z)=\exp\{[{\rm ad}({F}_2(Z))+{\rm ad}({F}_1(Z))]\}S(Z)=S({\bar Z}_2(Z))=
S({\bar Z}_1({\bar{\cal Z}}_2(Z))= \\
\label{S2}
&&=\exp\{{\rm ad}({{\cal F}}_2(Z))\}S_1(Z)
\end{eqnarray}
hold.
Using the chain rule and multiplication rule
for superdeterminants,  one obtains for the logarithm of
the superdeterminant of the anticanonical transformation (\ref{Z122})
\begin{eqnarray}
\nonumber
&&\ln\sDet[{\bar Z}^A_2(Z)\overleftarrow{\partial}_{B}]=
\ln\sDet\left[\left.{\bar Z}^A_1({\bar{\cal Z}}_2)
\overleftarrow{\partial}_{{\bar{\cal Z}}^C_2}\right|_{{\bar{\cal Z}}^C_2\rightarrow
{\bar{\cal Z}}^C_2(Z)}({\bar{\cal Z}}^C_2(Z)\overleftarrow{\partial}_{B}\right]= \\
\nonumber
&&=\ln\sDet\left[\left({\bar Z}^A_1({\bar{\cal Z}}_2)
\overleftarrow{\partial}_{{\bar{\cal Z}}^B_2}
\right)(Z)\right]+
\ln\sDet\left[{\bar{\cal Z}}^A_2(Z)\overleftarrow{\partial}_{B}\right]= \\
\label{sDetZ2}
&&=\exp\{{\rm ad}({\cal F}_2(Z))\}\ln\sDet\left[({\bar Z}^A_1(Z)
\overleftarrow{\partial}_{B})\right]+
\ln\sDet\left[{\bar{\cal Z}}^A_2(Z)\overleftarrow{\partial}_{B}\right].
\end{eqnarray}
Consider  the action $S'_2$
constructed from an action $S$ with the help of anticanonical
master transformation with
the generator functional $F_1+F_2$ (\ref{trF1F2}). We obtain
\beq
\label{Ss2}
S'_2(Z)=S_2(Z)-\frac{i\hbar}{2}\ln\sDet\left[{\bar Z}^A_2(Z)
\overleftarrow{\partial}_{B}\right]
\eeq
where $S_2(Z)$ is defined by the first equality in (\ref{S2}),
and ${\bar Z}^A_2$ is given by  the second equality in (\ref{trF1F2}). It follows
from (\ref{sDetZ2}) and (\ref{Ss2}) that
\beq
\nonumber
&&S'_2(Z)=\exp\{{\rm ad}({\cal F}_2(Z))\}\left(S_1(Z)-
\frac{i\hbar}{2}\ln\sDet\left[{\bar Z}^A_1(Z)
\overleftarrow{\partial}_{B}\right]\right)- \\
\nonumber
&&
-\frac{i\hbar}{2}\ln\sDet\left[{\bar{\cal Z}}^A_2(Z)
\overleftarrow{\partial}_{B}\right]= \\
\label{fSprime2}
&&=\exp\{{\rm ad}({\cal F}_2(Z))\}S'_1(Z)-
\frac{i\hbar}{2}\ln\sDet[{\bar{\cal Z}}^A_2(Z)\overleftarrow{\partial}_{B}],
\eeq
where $S'_1$ is master transformed action $S$ under anticanonical
transformation of variables $Z$ with the generator functional $F_1(Z)$,
and, as a result, $S'_2$
is presented as master transformed action $S'$ corresponding to
the anticanonical master transformation of $Z$ with generator functional
${\cal F}_2$,
i.e., in the form of successive anticanonical master transformations.
From (\ref{fSprime2}) we deduce the relations
\begin{eqnarray}
\nonumber
&&\exp\{-[F_2(z)+F_1(Z),\Delta]\}\exp\left\{\frac{i}{\hbar}S(Z)\right\}=
\exp\{-[{\cal F}_2(Z),\Delta]\}\exp\left\{\frac{i}{\hbar}S'_1(Z)\right\}= \\
&&=\exp\{-[{\cal F}_2(Z),\Delta]\}\exp \{-[F_1(Z),\Delta]\}
\exp\left\{\frac{i}{\hbar}S(Z)\right\}
\end{eqnarray}
being valid for arbitrary functional $S(Z)$.
The latter proves the relation (\ref{3.10}).

Finally, we give a proof of the relation
\beq
\frac{1}{2}\ln\sDet\left[{\bar Z}^A\overleftarrow{\partial}_{B}\right]=
-f({\rm ad}(F))\Delta F,\quad {\bar Z}^A=\exp\{{\rm ad}( F)\}Z^A,
\eeq
used in the representation of the master transformed actions (\ref{trAc})
and (\ref{Sprime2}). To do this, we introduce
a one-parameter family of anticanonical transformations
\beq
{\bar Z}^A(a)=\exp\{a{\rm ad}({F})\}Z^A,
\eeq
and the corresponding set of logarithm  of superdeterminants
\beq
D(a)=\ln\sDet\left[{\bar Z}^A(a)\overleftarrow{\partial}_{B}\right].
\eeq
Consider anticanonical transformations with the infinitesimal variation of the parameter $a$
\beq
{\bar Z}^A_2={\bar Z}^A(a+\delta a)=\exp\{(a+\delta a){\rm ad}({F})\}Z^A
\eeq
and functionals
\beq
D(a+\delta a)=\ln\sDet[{\bar Z}^A_2\overleftarrow{\partial}_{B}]=
\ln\sDet[{\bar Z}^A(a+\delta a)\overleftarrow{\partial}_{B}].
\eeq
Taking into account Eqs. (\ref{3.10}), (\ref{F1-F3}), (\ref{trF1F2})
and (\ref{trF3}),
we have the following identification
\beq
F_1=aF,\quad F_2=\delta a F,\quad {\cal F}_2=\delta a F
\eeq
and the representations up to the second order in $\delta a$
\beq
\label{D26}
\exp\{{\rm ad}({\cal F})_2\}=1+\delta a{\rm ad}(F)+O((\delta a)^2),\;
{\bar{\cal Z}}^A_2=Z^A+\delta aF\overleftarrow{\partial}_{C}E^{CA}+
O((\delta a)^2),
\eeq
\beq
\label{sDetZa}
\ln\sDet[{\bar{\cal Z}}^A_2\overleftarrow{\partial}_{B}]=\delta a
\sTr[F\overleftarrow{\partial}_{C}E^{CA}\overleftarrow{\partial}_{B}]+
O((\delta a)^2)=-2\delta a\Delta F+O((\delta a)^2).
\eeq
From (\ref{D26}), (\ref{sDetZa}) and (\ref{sDetZ2})
follows the differential equation for $D(a)$,
\beq
\partial_aD(a)={\rm ad}({F})D(a)-2\Delta F,\quad D(0)=0.
\eeq
Let us seek  a solution to this equation in the form
\beq
D(a)=\exp\{a{\rm ad}({F})\}D_1(a),\quad D_1(0)=0.
\eeq
Then we obtain
\beq
\partial_aD_1(a)=-2 \exp\{-a{\rm ad}({F})\}\Delta F, %\:\Rightarrow
\eeq
and
\beq
D_1(a)=-2a\exp\{-a{\rm ad}({F})\}f(a{\rm ad}({F}))\Delta F+C,\quad
C=D_1(0)=0.
\eeq
Finally, we find
\beq
D(a)=-2af(a{\rm ad}({F}))\Delta F,\;\;
\ln\sDet\left[{\bar Z}^A\overleftarrow{\partial}_{B}\right]
=D(1)=-2f({\rm ad}({F}))\Delta F.
\eeq

\appendix
\section*{Appendix E: Factorization of the Jacobian of
anticanonical transformation}
%\section{}
\setcounter{section}{5}
\renewcommand{\theequation}{\thesection.\arabic{equation}}
\setcounter{equation}{0}

For the sake of completeness of our study of anticanonical
transformations, let us present here a simple proof of the
factorization property of the grand Jacobian of an anticanonical
transformation within the field-antifield formalism \cite{BV,BV1}.
The result is known at least since the article \cite{BV2} of
Batalin and Vilkovisky, although the proof was omitted therein.

We will proceed with the use of antisymplectic  Darboux coordinates
$Z^{A}$ in the form of an explicit splitting  into  fields $\phi^{i}$
and antifields $\phi^*_{i}$,
 \beq
 \label{e1}
 Z^{A} = \{\phi^{i},\phi^*_{i}\},\quad \varepsilon(Z^A)=\varepsilon_A,\quad
\varepsilon(\phi^*_i)=\varepsilon(\phi^i)+1,
\eeq
so that
\beq
\label{e2}
 (Z^A,Z^B)=E^{AB},\quad \varepsilon(E^{AB})=\varepsilon_A+\varepsilon_B+1,
\eeq
where $E^{AB}$ is a constant invertible antisymplectic metric with the
following block structure
\beq
\label{e}
E^{AB}=\left(\begin{array}{cc}
0 & I\\
-I & 0\\
\end{array}\right)
\eeq
and antisymmetry property
\beq
\label{e3}
 E^{AB}=-(-1)^{(\varepsilon_A+1)(\varepsilon_B+1)}E^{BA}.
\eeq

Let $F = F(Z)$ be a fermion generator of an anticanonical
transformation,
\beq
\label{e4}
Z^{A}\;\rightarrow\; \bar{ Z }^{A}( t )=\exp\{t{\rm ad}( F)\}Z^A,
\quad \bar{ Z }^{A}( t = 0 ) = Z^{A},
\quad \bar{Z}^A=\{\Phi^i,\Phi^*_i\}.
\eeq
$\bar{Z}^A$ satisfies the Lie equation
\beq
\label{e5}
\frac{d}{dt}{\bar{ Z }}^{A}  =
( \bar{ F }, \bar{ Z }^{A} )_{ \bar{ Z }}, %(2)
\eeq
where $\bar{F} =  F( \bar{Z} ) = F(Z)$.

Let us consider the (grand) Jacobian, $J(t)$, of the anticanonical
transformation (\ref{e4})
\beq
\label{e6}
J(t)=\sDet\left[ {\bar Z}^A(t)\overleftarrow{\pa}_B\right],
\eeq
together with its logarithm
\beq
\label{e7}
\ln J(t)=\sTr \ln\left[ {\bar Z}^A(t)\overleftarrow{\pa}_B\right].
\eeq
By using (\ref{e5}) and the relations
\beq
\label{e8}
({\bar Z}^A\overleftarrow{\pa}_C)(Z^C\overleftarrow{\pa}_{{\bar B}})=
\delta^A_B,\quad
(Z^A\overleftarrow{\pa}_{{\bar C}})({\bar Z}^C\overleftarrow{\pa}_B)=\delta^A_B,
\eeq
which are valid for any invertible transformation $Z^{A}\;\rightarrow \;
\bar{Z}^{A}$,
together with the formula for a $\delta$-variation,
\beq
\delta \sTr\ln M=\sTr M^{-1}\delta M, \quad (M^{-1})^A_CM^C_B=\delta^A_B,
\eeq
we derive the equation for $\ln J$
\beq
\nonumber
&&\frac{d}{dt}\ln J=\sTr \left[(Z^A\overleftarrow{\pa}_{\bar{C}})
\frac{d}{dt} (\bar{Z}^C\overleftarrow{\pa}_B)\right]=(-1)^{\varepsilon_A}
(Z^A\overleftarrow{\pa}_{\bar{C}})
(\dot{\bar{Z}}^C\overleftarrow{\pa}_A)=\\
\nonumber
&&=(-1)^{\varepsilon_A}(Z^A\overleftarrow{\pa}_{\bar{C}})
((\bar{F},{\bar Z}^C)\overleftarrow{\pa}_A)
=-(-1)^{\varepsilon_C}(\overrightarrow{\pa}_{\bar{C}}Z^A)
\overrightarrow{\pa}_A({\bar Z}^C,{\bar F})=\\
\label{ee9}
&&=-(-1)^{\varepsilon_C}\overrightarrow{\pa}_{\bar{C}}(\bar{Z}^C,\bar{F})=
-2\bar{\Delta}\bar{F},
\eeq
where the operators $\Delta$, $\bar{\Delta}$ are
defined\footnote{Notice that in (\ref{ee9}) we mean just the second
equality (\ref{e9}) so as to define the
transformed operator $\bar{\Delta}$. That definition
is maintained by the two following
motivations : it respects both the
nilpotency property and the multiplicative composition $\overline{\Delta G} =
\bar{\Delta} \bar{G}$, $\bar{G} = G(\bar{Z}$), with
arbitrary function $G = G(Z)$.} by
\beq
\nonumber
&&\Delta =\Delta_Z= \frac{1}{2} (-1)^{ \varepsilon_{A} } \pa_{A} ( Z^{A}, ... )=
\frac{1}{2} (-1)^{ \varepsilon_{A} } \pa_{A}E^{AB}\pa_B,\\
\label{e9}
&&\bar{\Delta}=\Delta_{\bar{Z}}=
 \exp\{ {\rm ad}(t F) \} \Delta \exp\{ {\rm ad}( - t F) \}.  %   (3)
\eeq
Here $\pa_A$ and $\pa_{{\bar A}}$ denotes  partial $Z^A$- and
$\bar{Z}^A$-derivative, respectively.

Now, let $J_{\phi}$ be  the Jacobian in the sector of fields,
\beq
J_{ \phi }(t)= \sDet
\left[\Phi^i(t,\phi,\phi^*)\overleftarrow{\pa}_j\right],
\eeq
together with its logarithm,
\beq \ln J_{\phi}(t)=
\sTr\ln\left[\Phi^i(t,\phi,\phi^*)\overleftarrow{\pa}_j\right],
\eeq
where $\pa_i$ denotes  partial $\phi^i$-derivative.  In what follows
below, the symbols $\pa_{\bar{k}}$ and $\pa^{*\bar{k}}$, with barred indices,
will be used to denote  partial $\Phi^k$- and $\Phi^*_k$-derivatives,
respectively. To get the $t$-derivative of $\ln J_{\phi}$, one needs the
inverse to the matrix $\Phi^i\overleftarrow{\pa}_j$.

Let us consider  an anticanonical  transformation in the sector of
fields,
\beq
\label{ee1}
\phi^i\;\rightarrow\;\Phi^i=\Phi^i(t,\phi,\phi^*).
\eeq
Let us resolve that equation for initial fields $\phi^{i}$, with $t$ and
$\phi^*_{i}$ kept fixed,
\beq
\label{ee2}
\phi^i=\phi^i(t,\Phi,\phi^*),
\eeq
so that
\beq
\label{ee3}
\phi^i(t,\Phi(t,\phi,\phi^*),\phi^*)\equiv\phi^i.
\eeq
It follows from (\ref{ee3}) that the relation
\beq
\left(\phi^i(t,\Phi,\phi^*) \overleftarrow{\pa}_{\bar{k}}\right)
\left(\Phi^k(t,\phi,\phi^*)\overleftarrow{\pa}_j\right)=\delta^i_j,
\eeq
holds, because  the initial fields $\phi^{i}$ are inverse
functions to $\Phi^i(t, \phi,\phi^*)$ at the fixed
values of $t$ and $\phi^*_{i}$.  From now on, the variables
$\Phi^{i}, \Phi^*_{i}$  are considered
as  functions of $t$, $\phi^{i}$, $\phi^*_{i}$, while the fields $\phi^{i}$
are functions of $t$, $\Phi^{i}$, $\phi^*_{i}$,
so that the short notation will be used naturally,
\beq
\phi^i(t,\Phi,\phi^*)=\phi^i,\quad \Phi^i(t,\phi,\phi^*)=\Phi^i,\quad
\Phi^*_i(t,\phi,\phi^*)=\Phi^*_i.
\eeq

Now, we derive the following equation for  $\ln J_{\phi}$
\beq
\frac{d}{dt}\ln J_{\phi}=
-\bar{\Delta}\bar{F}-\frac{1}{2}(-1)^{\varepsilon_k}\bar{F}
\overleftarrow{\pa}^{*\bar{k}}\overleftarrow{\pa}^{*\bar{m}}\left[
(\Phi^*_m\overleftarrow{\pa}_i)
(\phi^i\overleftarrow{\pa}_{\bar{k}})-
(k\;\leftrightarrow\;m)(-1)^{\varepsilon_k\varepsilon_m}\right].
\eeq
In turn, let us consider the quantity,
\beq
\label{ee21}
T_{jk}=(\Phi^*_j\overleftarrow{\pa}_i)(\phi^i\overleftarrow{\pa}_{\bar{k}})-
(\Phi^*_k\overleftarrow{\pa}_i)(\phi^i\overleftarrow{\pa}_{\bar{j}})
(-1)^{\varepsilon_j\varepsilon_k}.
\eeq
Then, by multiplying (\ref{ee21}) subsequently from the right by the two
Jacobi matrices accompanied with a special sign factor, we have,
\beq
T_{jk}(\Phi^k\overleftarrow{\pa}_l)(\Phi^j\overleftarrow{\pa}_m)
(-1)^{\varepsilon_j\varepsilon_l}=(\overrightarrow{\pa}_l\Phi^*_j)
(\Phi^j\overleftarrow{\pa}_m)-(m\;\leftrightarrow\;l)
(-1)^{\varepsilon_m\varepsilon_l}.
\eeq
The latter can be rewritten in the form,
\beq
\label{ee23}
T_{jk}(\Phi^k\overleftarrow{\pa}_l)(\Phi^j\overleftarrow{\pa}_m)
(-1)^{\varepsilon_j\varepsilon_l}=(\overrightarrow{\pa}_l\Phi^*_j)
(\Phi^j\overleftarrow{\pa}_m)-(\overrightarrow{\pa}_l\Phi^j)
(\Phi^*_j\overleftarrow{\pa}_m)
\eeq
Now, let us introduce the quantity,
\beq
\label{ee24}
L_{AB}=(\overrightarrow{\pa}_A {\bar Z}^C)E_{CD}(\bar{Z}^D\overleftarrow{\pa}_B)
\eeq
where $E_{AB}$ is the inverse to $E^{AB}$, with the following block
structure,
\beq
\label{e0}
E_{AB}=\left(\begin{array}{cc}
0 & -I\\
I & 0\\
\end{array}\right),\quad \varepsilon(E_{AB})=\varepsilon_A+\varepsilon_B+1
\eeq
and the antisymmetry property
\beq
E_{AB}=-(-1)^{\varepsilon_A\varepsilon_B}E_{BA}.
\eeq
Notice that the field-field  components of (\ref{ee24}),
\beq
L_{ij}=(\overrightarrow{\pa}_i\Phi^*_k)
(\Phi^k\overleftarrow{\pa}_j)-(\overrightarrow{\pa}_j\Phi^k)
(\Phi^*_k\overleftarrow{\pa}_i),
\eeq
do coincide with (\ref{ee23}).
By taking the relation
\beq
(Z^A,Z^B)_{\bar{Z}}=(Z^A\overleftarrow{\pa}_{\bar{C}})E^{CD}
(\overrightarrow{\pa}_{\bar{D}}Z^B)=E^{AB}
\eeq
into account, we have
\beq
\nonumber
&&E^{AC}L_{CB}=(Z^A,Z^C)_{\bar{Z}}L_{CB}=(Z^A\overleftarrow{\pa}_{\bar{C}})E^{CD}
(\overrightarrow{\pa}_{\bar{D}}Z^E)
(\overrightarrow{\pa}_E {\bar Z}^F)E_{FG}(\bar{Z}^G\overleftarrow{\pa}_B)=\\
&&=(Z^A\overleftarrow{\pa}_{\bar{C}})E^{CD}\delta^F_D
E_{FG}(\bar{Z}^G\overleftarrow{\pa}_B)=(Z^A\overleftarrow{\pa}_{\bar{C}})
(\bar{Z}^C\overleftarrow{\pa}_B)=\delta^A_B.
\eeq
The latter implies\footnote{The same result follows
via $t$-differentiation of (\ref{ee21})
and the use of the Lie equation (\ref{e5}).}
\beq
L_{AB}=E_{AB},\quad L_{ij}=0.
\eeq
Thus, we obtain the equation for the Jacobian $J_{\phi}$ in the sector
of fields,
\beq
\label{e10}
\frac{d}{dt}{ \ln J_{ \phi } } = -  \bar{ \Delta}  \bar{ F }.   % (4)
\eeq

In the same way, we derive the equation
\beq
\label{e11}
\frac{d}{dt}{ \ln J_{ \phi^* } } = - \bar{ \Delta} \bar{ F }   % (5)
\eeq
for the Jacobian $J_{\phi*}$ in the sector of antifields,
\beq
J_{ \phi^* }(t)=\sDet\left[\Phi^*_i(t,\phi,\phi^*)\overleftarrow{\pa}^{*j}\right],
\quad \ln J_{ \phi^* }(t)=
\sTr\ln\left[\Phi^*_i(t,\phi,\phi^*)\overleftarrow{\pa}^{*j}\right].
\eeq

It follows from (\ref{ee9}), the initial data (\ref{e4}),
(\ref{e10}) and (\ref{e11}), that
\beq
\label{E6}
J_{ \phi }  =  J_{ \phi^* }  =  J^{1/2},     %      (6)
\eeq
and finally, the factorization property,
\beq
\label{E7}
J  =  J_{ \phi } J_{ \phi^* }.    %   (7)
\eeq

It seems to be rather useful to mention here the main properties of the
grand Jacobian $J$ of anticanonical transformations,
within the field-antifield formalism. Let
$Z^{A}\;\rightarrow \; \bar{Z}^{A}$
be an anticanonical transformation with a fermion generator $F$.
Consider Eq. (\ref{ee9}) as rewritten in the form
\beq
\label{eq2}
\frac{d}{dt}\ln J^{ 1/2 } = - \bar{\Delta} \bar{F},       %    (2)
\eeq
where the $\bar{\Delta}$-operator is defined in (\ref{e9}).
A formal solution to (\ref{eq2}) has the form \beq
\label{eq4}
\ln J^{ 1/2 }= - [ ( \exp\{{\rm ad}( t F ) \} - 1 ) /{\rm ad}(F)] \Delta F.%   (4)
\eeq
It follows immediately from (\ref{e9})
\beq
\nonumber
&&\frac{d}{dt}\bar{\Delta}=\exp\{ {\rm ad}( t F) \}[{\rm ad}( F),\Delta]
\exp\{{\rm ad}( - t F ) \} =\\
&&={\rm ad}( - \exp\{ {\rm ad}( t F ) \} \Delta F  )=\frac{d}{dt}
{\rm ad}( \ln J^{ 1/2 } ),
\eeq
which implies\footnote{The same result follows from (\ref{e9}) and the use of
the anticanonical invariance of $E^{AB}$.}
\beq
\label{eq6}
\bar{\Delta} = \Delta + {\rm ad}(\ln J^{ 1/2 } ).    %   (6)
\eeq
That is just the transformation property of the $\Delta$ operator under
anticanonical transformation.
Further, it  follows from (\ref{eq2})
\beq
\label{eq7}
\bar{\Delta}  \frac{d}{dt}  (\ln J^{ 1/2 } ) = 0.         %  (7)
\eeq
By substituting (\ref{eq6}), we get
\beq
\label{eq8}
 \frac{d}{dt} \left[  \frac{1}{2} (   \ln J^{ 1/2 },   \ln J^{ 1/2 } )
  + \Delta \ln J^{ 1/2 }  \right] =0, %(8)
\eeq
which implies
\beq
\label{eq9}
\Delta \exp\left\{ \ln J^{ 1/2 } \right\} = \Delta ( J^{ 1/2 } )  = 0.   %   (9)
\eeq
That is just the antisymplectic counterpart to the Hamiltonian Liouville
theorem \cite{BV2,BBD}.
\\

\end{document}